\documentclass[sigconf, authorversion]{acmart}
\AtBeginDocument{%
  }

\usepackage{soul}
\usepackage{subcaption} %
\usepackage{enumitem} %

\copyrightyear{2026}
\acmYear{2026}
\setcopyright{cc}
\setcctype{by}
\acmConference[VRST '26]{32nd ACM Symposium on Virtual Reality Software and Technology}{November 16--18, 2026}{Sendai, Japan}
\acmBooktitle{32nd ACM Symposium on Virtual Reality Software and Technology (VRST '26), November 16--18, 2026, Sendai, Japan}
\acmDOI{10.1145/3822517.3848657}
\acmISBN{979-8-4007-2811-2/2026/11}

\begin{document}

\title{Steering Versus Teleporting in Mobile Virtual Reality}

\author{Kristen Grinyer}
\email{kristengrinyer@cmail.carleton.ca}
\orcid{0009-0005-4098-3165}
\affiliation{%
  \institution{Carleton University}
  \city{Ottawa}
  \state{ON}
  \country{Canada}
}
\author{Daniel Zielasko}
\email{daniel.zielasko@rwth-aachen.de}
\orcid{0000-0003-3451-4977}
\affiliation{%
  \institution{Technical University of Denmark}
  \city{Ballerup}
  \state{}
  \country{Denmark}
}
\author{Robert J. Teather}
\email{rob.teather@monash.edu}
\orcid{0009-0007-4572-1820}
\affiliation{%
  \institution{Monash University}
  \city{Melbourne}
  \state{Victoria}
  \country{Australia}
}

\renewcommand{\shortauthors}{Grinyer et al.}

\begin{abstract}
Mobile virtual reality (MVR) provides low‑cost access to extended reality (XR), but its limited input restricts use of common locomotion techniques such as head-decoupled and velocity-controlled steering. Using a low‑cost controller with an audio-based button, we compared gaze‑ and controller‑directed steering and teleportation in MVR. We included a controller pitch-based speed-control technique enabling continuous head-decoupled steering. Performance and participant feedback indicate that gaze was better suited to teleportation, whereas controller pointing better supports steering in primed search. We compare with similar techniques in standard VR and derive design considerations for low-cost, narrow-FOV hardware, demonstrating the potential to support varied travel techniques and velocity control in low-fidelity XR.
\end{abstract}


\begin{teaserfigure}
    \centering
    \begin{subfigure}{0.33\linewidth}
    \includegraphics[width=.99\textwidth]{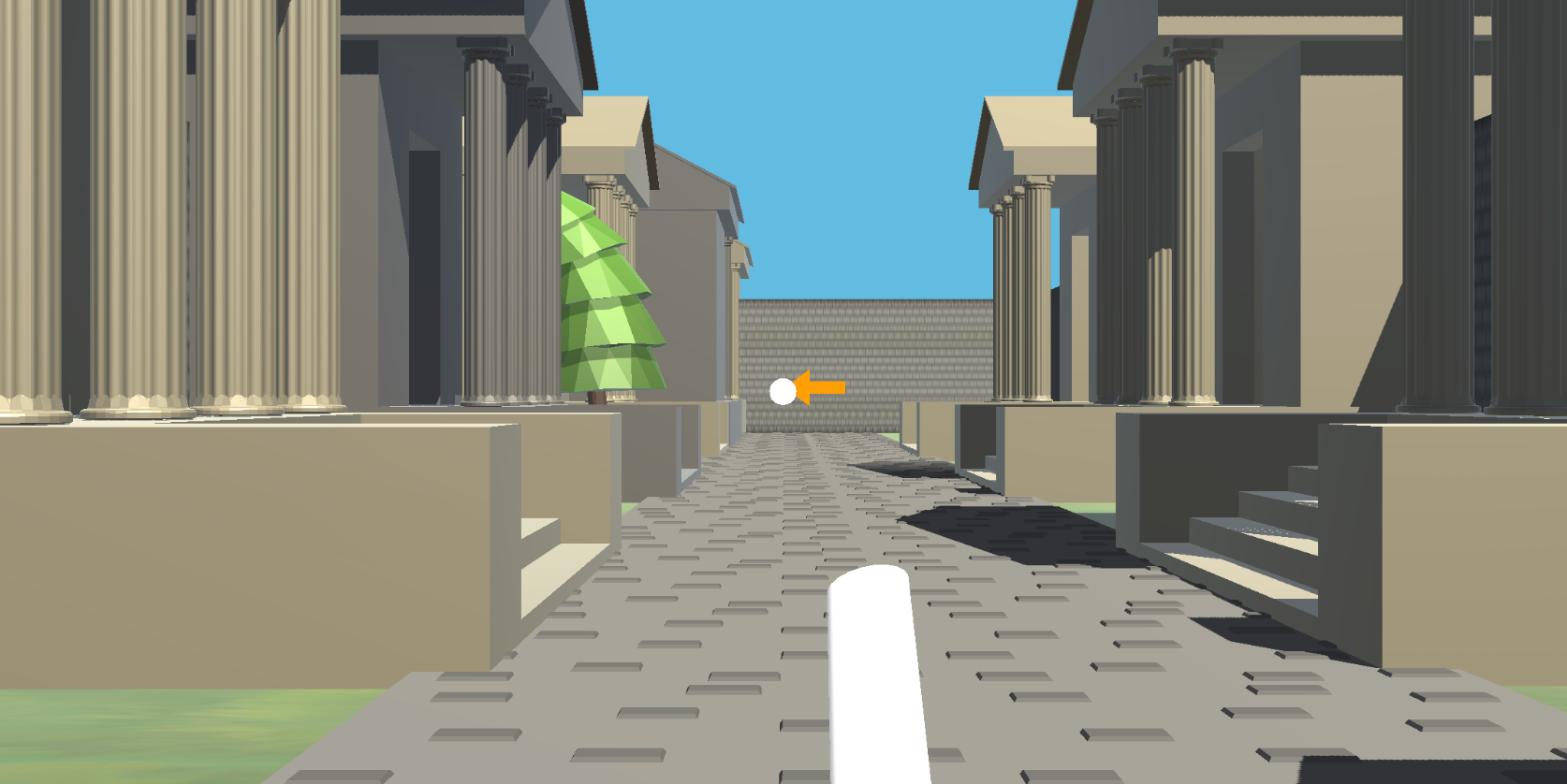}
    \caption{Gaze- and controller-directed steering}
    \label{fig_controller-steering}
    \Description{VR user view of a virtual pathway; user is holding and pointing a white virtual controller ahead.}
    \end{subfigure}
    \hfill
    \begin{subfigure}{0.33\linewidth}
    \includegraphics[width=.99\textwidth]{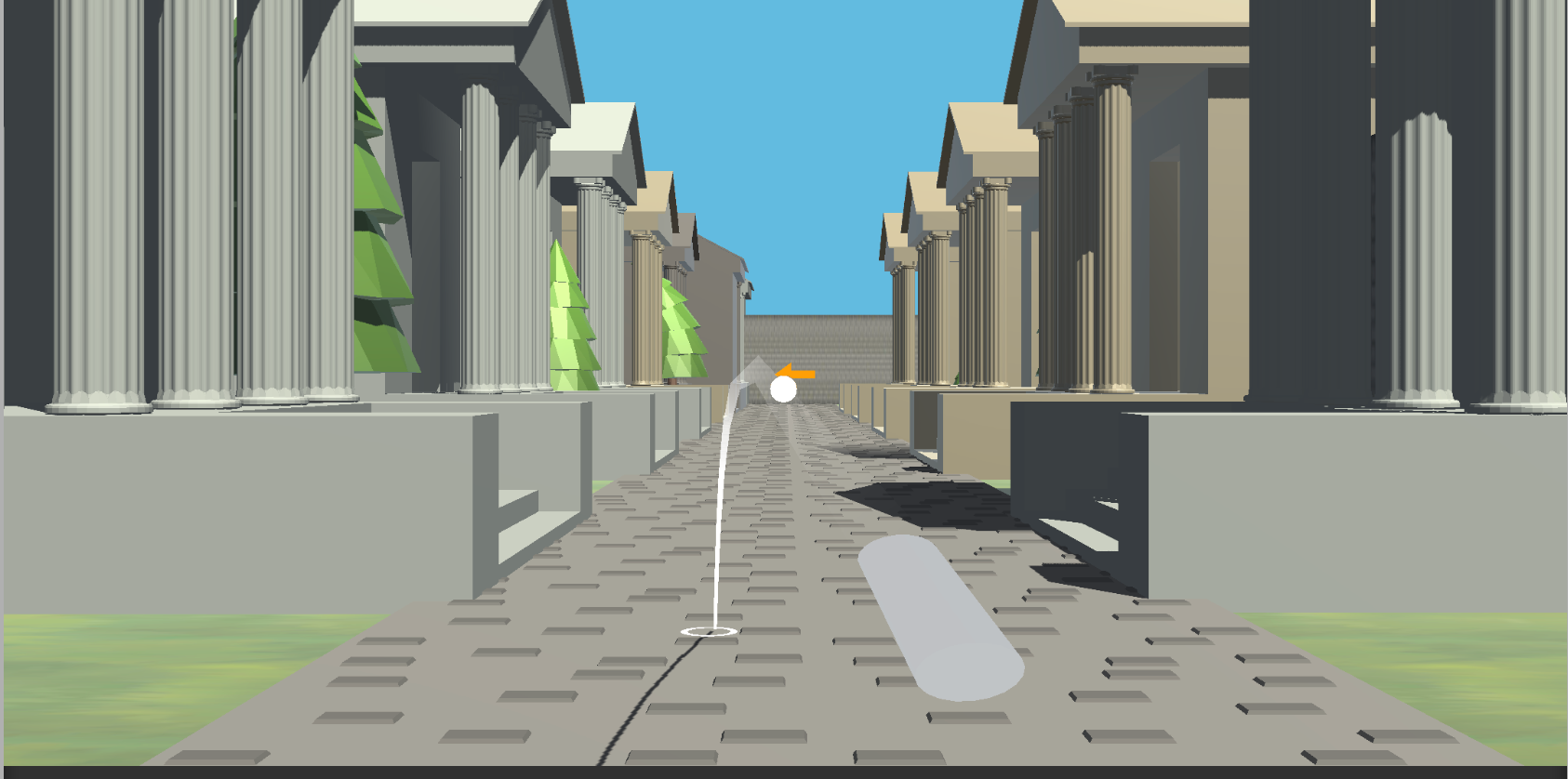}
    \caption{Controller-directed teleportation}
    \label{fig_controller-teleport}
    \Description{VR user view of a virtual pathway; user is holding a white virtual controller with a projectile arc shooting out and intersecting the ground. A flat ring indicates the teleport position.}
    \end{subfigure}
    \hfill
    \begin{subfigure}{0.33\linewidth}
    \includegraphics[width=.99\textwidth]{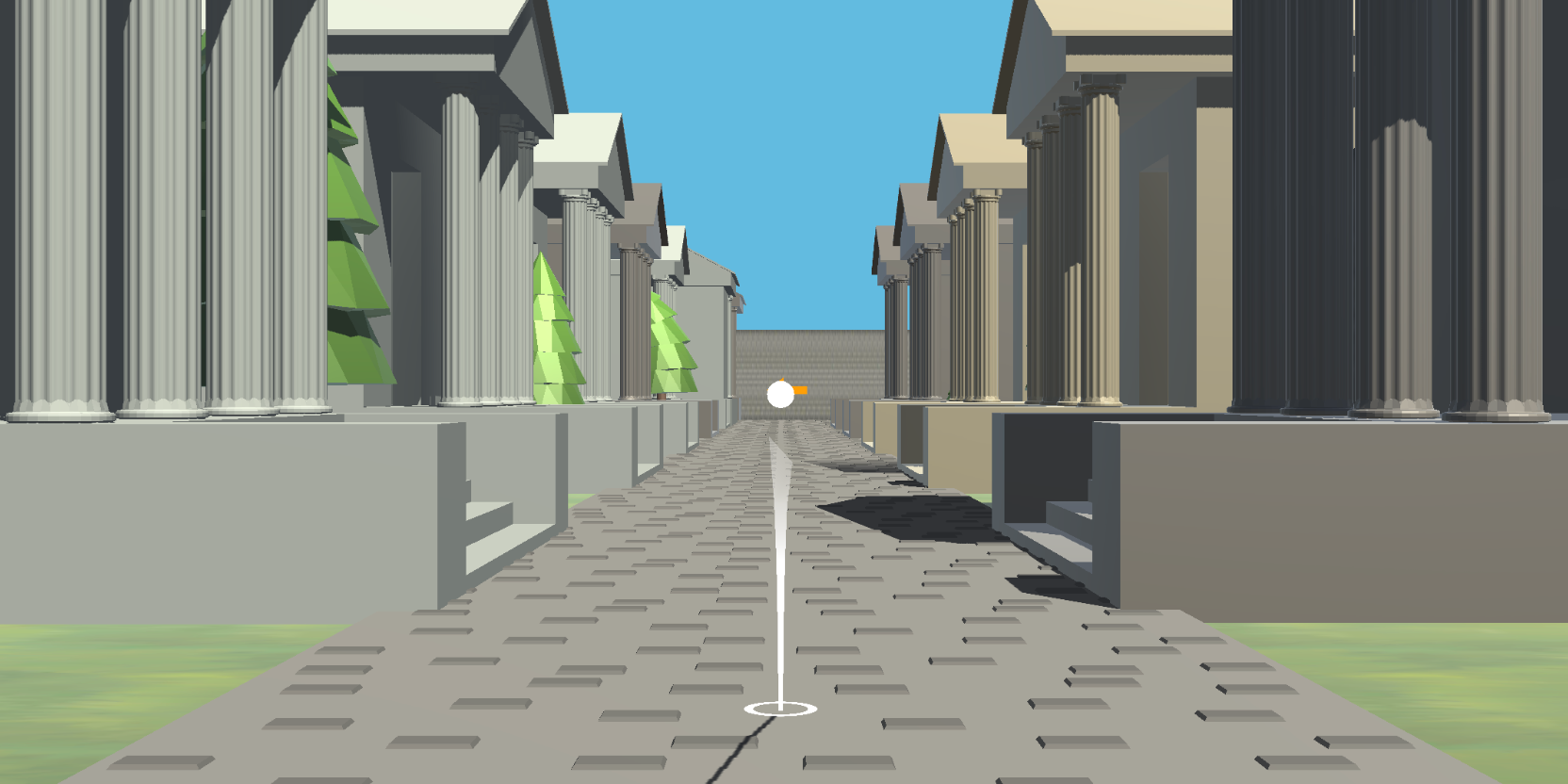}
    \caption{Gaze-directed teleportation}
    \label{fig_gaze-teleport}
    \Description{VR user view of a virtual pathway; a white projectile arc extending from the user's head position intersects the ground and a flat ring indicates the teleport position.}
    \end{subfigure}  
    \caption{Travel techniques from user's point-of-view, showing experiment environment and virtual controller.}
    \Description{Three images showing VR user view of virtual environment including a cobblestone pathway with grass and temple buildings on either side.}
    \label{fig:conditions}
\end{teaserfigure}

\maketitle

\section{Introduction}
Extended reality (XR), encompassing augmented, mixed, and virtual reality (AR, MR, and VR), is increasingly integrated into personal and professional life. Education, healthcare, and retail markets are driving adoption \cite{MarketReport2026}, while VR is now also used in hiring \cite{Beti2019}. Consequently, XR access and skills are increasingly important for equitable participation across socioeconomic backgrounds. This raises concerns about the digital divide:  disparities in 1) technology access, 2) usage behaviour and skills (how people use technology \cite{Bonfadelli2002, Hargittai2008} and capacity to do so \cite{Litt2013}), and 3) the ability to derive benefits from tech access \cite{Ragnedda2018, Ragnedda2017, Lythreatis2022}. Current XR technologies may widen access disparities, exacerbating the other two levels.

Researchers and practitioners have a societal obligation to support equitable access and to continue developing cost-friendly, lower-fidelity XR hardware. Mobile VR (MVR) uses a smartphone as the primary hardware and provides a compelling and inexpensive alternative (\$10–60 USD) to standalone VR head-mounted displays (HMD) (\$300–3500 USD). Smartphones are universally adopted regardless of income or education level \cite{PewResearch}; MVR can leverage this mass adoption to improve access to immersive experiences. However, to date, MVR lacks the capability to run the range of applications or deliver the advantages of immersive VR.  

Current MVR devices typically provide only primitive, if any, input \cite{Grinyer2026-Review}. While MVR supports 3-degrees-of-freedom (DOF) head-tracking, discrete input is inconsistent with capabilities ranging from HMD-embedded buttons to untracked Bluetooth controllers, or nothing at all \cite{Gao2025}. Few, if any, MVR devices include out-of-the-box 6DOF tracked controllers, despite these being universally provided with all commercial standalone VR HMDs. Some smartphone-powered MR headsets include tracked input devices \cite{ZapBox}, but do not support fully virtual environments and impose substantial barriers to access (e.g., limited OS support and regional availability). These severely limited input capabilities mean MVR cannot support comparable immersive experiences or provide the 3D interaction required of high-end VR experiences %
\cite{Argelaguet2013,LaViolaJr2017,Bowman1998Travel,Yu2025}.

Prior MVR interaction research examined fundamental 3D tasks \cite{Bowman2001} including selection (target acquisition) \cite{Hirzle2018, Grinyer2023, Dani2018, Zhang2017, Grinyer-TVCG, Mohr2019, Silva2024} and travel (viewpoint control) \cite{Caggianese2015,Amaral2024, Hedlund2023, Hanson2019, Shanmugam2017, Tregillus2017, Ang2019, Wang2019, Hudk2020, Berriel2018}. Common travel techniques, like teleportation and steering, typically require at least one tracked point and discrete input. Due to MVR's hardware constraints, travel often uses gaze-directed steering with start/stop controls \cite{Powell2017} or gaze-directed teleportation using dwell confirmation \cite{Bothen2018,Dresel2021,Bozgeyikli2016}. MVR travel research has studied walk-in-place (WIP) \cite{Hedlund2023,Hanson2019,Shanmugam2017,Tregillus2016,Tregillus2017,Ang2019,Wang2019} and real walking \cite{Wang2019,Hudk2020,Berriel2018}. However, real walking is constrained by physical space, and WIP is inefficient for large virtual environments (VEs). %
Virtual movement techniques, like steering, avoid spatial limitations, but may induce cybersickness due to visual-vestibular conflicts \cite{Ang2023Cybersickness}. Teleportation mitigates both issues but remains underexplored in MVR \cite{Grinyer2026-Review}. Although prior studies evaluated teleportation \cite{Rupp2026,Prithul2021}, they used VR hardware that was not subject to MVR's interaction constraints. We therefore compare travel techniques on low‑fidelity XR hardware, where interaction limitations may affect usability and UX, and findings from standalone VR may not generalize.

Our work is motivated by the research question: \textit{Can low-fidelity hardware support effective and usable travel in mobile VR?} We studied travel via teleportation and steering using our previously evaluated low-fidelity controller for MVR \cite{Grinyer2023,Grinyer-TVCG,Grinyer-GI26}. Consistent with MVR's affordability, the Cardboard Controller is tracked in 6DOF by the smartphone camera and uses an acoustically sensed button \cite{Grinyer-GI26}. We evaluated free-range teleportation and velocity-controlled steering decoupled from view direction. Despite their benefits head-decoupled steering \cite{Christou2016CAVE,ZielaskoTestBed, Hegde2016} and velocity control \cite{Powell2017,Rahimian2021, Argelaguet2014,Zielasko2025Bimanual}, lack standard MVR implementations and remain uncommon in MVR research \cite{Grinyer2026-Review}, while %
velocity control is underexplored in VR locomotion in general \cite{Martinez2022}. We compared four combinations of controller- and gaze-directed pointing across steering and teleportation (see Figure \ref{fig:conditions}). Our contributions include:
\begin{itemize}[leftmargin=*]
\item Evaluation of a motion-tracked cardboard controller and acoustically sensed activation for supporting teleportation and viewpoint-decoupled steering techniques in MVR.
\item A comparison of steering and teleportation, assessing whether their known advantages generalize to large environments using low‑cost hardware.
\item Adaptation and evaluation of controller-pitch mapping for omnidirectional speed control on low-cost MVR hardware.
\end{itemize}

\section{Related Work}
Travel, or locomotion, is a fundamental VR interaction enabling viewpoint control \cite{Bowman1997,LaViolaJr2017,Powell2017}. It comprises three components: direction/target selection, velocity/acceleration, and input (how travel starts/stops) \cite{Bowman1997,Zielasko2025Bimanual}. Common approaches include walking, steering, and selection-based travel \cite{LaViolaJr2017,Prinz2023}. Walking techniques are well-studied in MVR \cite{Grinyer2026-Review}, but impractical for large VEs.

\subsection{Steering-Based Travel} \label{sec: related-work-steering}
Steering techniques commonly use a joystick or gaze direction for locomotion unconstrained by real-world space \cite{LaViolaJr2017}, but can induce cybersickness due to mismatched visual and vestibular cues \cite{Ang2023Cybersickness}. Because most off-the-shelf MVR hardware lacks continuous input, steering techniques typically omit speed control and rely on automated motion or start/stop input \cite{Powell2017, Grinyer2026-Review}. Automated steering is less preferred since it reduces user control \cite{Powell2017,Bothen2018}. %
Past work explored decoupling viewpoint and steering control using smartwatch-detected wrist motions \cite{Amaral2024, Silva2024}, though this was not formally evaluated.

Powell et al. \cite{Powell2017} evaluated steering in Google Cardboard across three control levels: forward movement, forward movement with start/stop, and forward/backward movement with start/stop using a Bluetooth controller. Although no technique was ideal, participants preferred the controller for ease of use. Leaning-based interfaces similarly map body or head tilt to velocity, providing more natural self-motion cues and better usability than joysticks \cite{Hashemian2023}. Building on this work, our controller-based technique uses intuitive tilt-based speed control, retaining performance and preference benefits associated with motion-tracked controls over joysticks \cite{Buttussi2019, ZielaskoTestBed, Hashemian2023, Harris2014, Marchal2011}, and the advantages of head-decoupled movement \cite{Christou2016CAVE,ZielaskoTestBed, Hegde2016}. We introduce omnidirectional speed control---an essential travel component \cite{Bowman1997} absent from MVR steering techniques \cite{Grinyer2026-Review}. By evaluating head-decoupled steering with full X-Z axes movement, start/stop, and speed control, we present a comprehensive solution addressing MVR user-control limitations.

\subsection{Selection-Based Travel}
With selection-based travel, users select a target point or path \cite{LaViolaJr2017,Prinz2023} and the viewpoint then moves to the destination, either instantly (teleportation), or sometimes through animated transitions \cite{Feld2025,Wolwer2025}. %
Teleportation is efficient, easy to use, and avoids both cybersickness and environment size constraints, though it may disorient users \cite{LaViolaJr2017,Ang2023Cybersickness,Bowman1997}. Teleportation confirmation can employ either an HMD button or dwell timers by holding head/controller direction stationary for 2–3 seconds \cite{Bothen2018,Dresel2021,Bozgeyikli2016}. Bozgeyikli et al. \cite{Bozgeyikli2016} used a controller with dwell and found no difference in travel time compared to button confirmation. Comparing dwell, voice, and eye-blink to finger-pinch \cite{Narbayev2025} or foot-stomp \cite{Spurgeon2018} for confirmation, finger-pinch was fastest and offered better control \cite{Narbayev2025}. Voice was slower due to speaking time \cite{Narbayev2025, Spurgeon2018}, while foot-stomp was least accurate and required twice as many teleports \cite{Spurgeon2018}. Compared to controller-based methods, blink and dwell were adequate and preferred by users, though controllers offered slightly better performance than all hands-free methods. These findings support further exploration of controller-based teleportation in MVR.

\subsection{Comparing Teleportation and Steering}
Studies consistently find that teleportation is faster and yields less cybersickness than steering \cite{Buttussi2019, Dylong2026, Clifton2020, Christou2017}, while steering enhances presence \cite{Clifton2020}. Christou et al. compared gaze- and controller-directed steering in CAVE and HTC VIVE environments \cite{ Christou2016CAVE, Christou2017}, including teleportation in the latter. In a token-collection task, teleportation yielded more missed tokens and environmental details \cite{Christou2017}, while head-decoupled steering improved performance \cite{Christou2016CAVE}. The authors observed that participant strategies---and thus travel technique effectiveness---are impacted by field of view (FOV) size: gaze-directed travel is best for small-FOV systems, whereas controller pointing better suits CAVE environments \cite{Christou2017}.

Noting past findings and the unique constraints of MVR devices (e.g., input capabilities, field of view, etc.), we argue for further investigation of locomotion techniques in MVR. No prior work has examined how travel technique affects MVR travel performance, user strategy, and preference. Our study explores which technique best navigates large VEs in low-fidelity immersive displays.

\subsection{Locomotion Performance Evaluation}
Surveys of VR locomotion evaluation \cite{Cannavo2020,Martinez2022} report continued reliance on cumulative performance metrics, such as task completion time, total distance travelled, and collision count \cite{Mousas2021, Garcia2024, Tregillus2016}. These indicate general efficiency but provide limited insight into travel trajectories or path deviation. This motivated trajectory-based evaluations that compared average trajectory curvature and spatio-temporal path deviation relative to reference paths \cite{Mousas2021, Terziman2011}. However, such approaches often require multiple complementary measures to quantify performance or fail to capture nuanced differences in locomotion behaviour. %
Meanwhile, waypoint-based tasks typically use waypoints that are not equally spaced apart \cite{Kontio2023,Prithul2022,Habgood2018}, which makes them less suitable for trajectory-based analyses. 

Locomotion literature has recently adopted human-motor models such as Fitts’ law \cite{MacKenzie2018} and the Steering law \cite{Accot1997}. These models have been used to study teleportation \cite{Sindhupath2024, Rupp2026} and VR walking \cite{Monteiro2018}, suggesting that trajectory-based approaches can characterize both discrete and continuous VR travel, but they remain underexplored. 
Recent work proposed novel testbeds comparing locomotion techniques because no standard approach exists \cite{Sarupuri,Cannavo2020}. Sarupuri et al. developed the Locomotion Usability Test Environment (LUTE) for travel task evaluation supporting personalized parameters \cite{Sarupuri}. Cannavò et al. introduced the Locomotion Evaluation Testbed for VR (LET-VR), combining a custom experimental measurement protocol with weighted scoring to rank locomotion techniques \cite{Cannavo2020}.

Following this approach, we sample participant position over time to evaluate user trajectories and accuracy more finely than cumulative measures (i.e., completion time, total distance travelled). Unlike waypoint-based approaches evaluating deviation at predefined locations \cite{Lee2009}, our method enables fine-grained analysis of steering behaviour and path conformity over time.

A 2022 review on VR locomotion \cite{Martinez2022} found hand-directed steering and velocity control are largely overlooked. While point-and-teleport is common in standalone VR, a 2026 review of MVR interaction found teleportation and selection-based travel neglected \cite{Grinyer2026-Review}. It is thus unclear whether the known advantages of teleportation in higher-fidelity VR hold for MVR. We address these gaps by investigating a speed-control technique for controller-directed steering and comparing it to teleportation in MVR.

\section{System Design}
\begin{figure}[!t]
\centering
    \includegraphics[width=.9\linewidth]{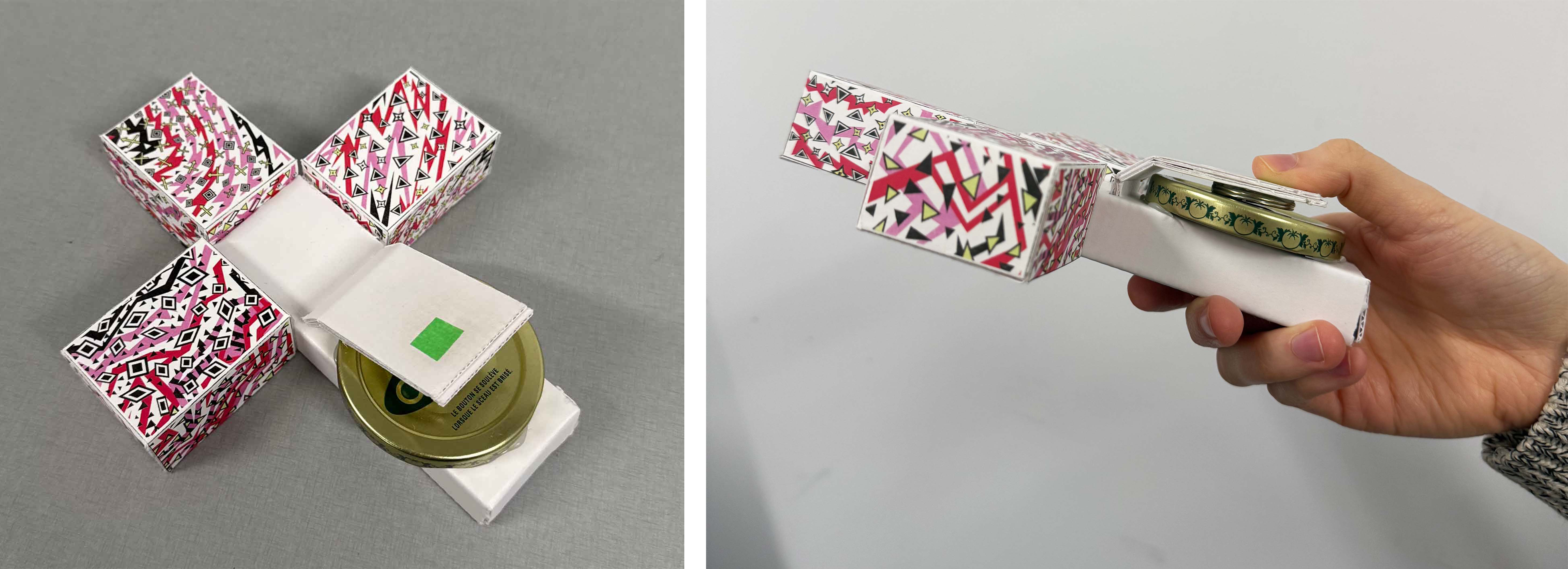}
    \caption{Controller apparatus (left) and button press (right).}
    \label{fig_controller}
    \Description{(Left) Cardboard folded into a rectangular handle. Three cubic paper markers with colourful patterns are attached to the top left, front, and right sides of the handle. A jar lid is fastened to the top of the handle, and a second cardboard piece covers it. (Right) Side view of user holding controller and pressing down on jar lid using second cardboard piece.}
\end{figure}

We used our low-fidelity Cardboard Controller \cite{Grinyer2023,Grinyer-TVCG} with three paper markers tracked in 6DOF by the smartphone camera (Figure \ref{fig_controller}). Controller orientation was updated each frame from one marker's quaternion, retaining the prior frame's source or the top marker to compensate for tracking loss. We ignored yaw changes exceeding 12° to prevent jitter and inversions exceeding 90° producing ergonomically awkward movements.

Input activation in MVR commonly uses HMD buttons or dwell \cite{Grinyer2026-Review}. However, dwell is poorly suited to teleportation; avoiding the "Midas Touch" effect (accidental activation), which can disorient users \cite{Jacob1990}, requires preset teleport points that restrict locomotion freedom. The Google Cardboard button is also unsuitable because: 1) it is not universally available, 2) its position causes arm fatigue, and 3) registration errors and headset movement when pressed reduce accuracy (i.e., the "Heisenberg" effect \cite{Wolf2020}) \cite{Powell2017,Bothen2018,Smus2015}. The cardboard controller instead uses an acoustically sensed, top-mounted jar lid as a button \cite{Grinyer-GI26}. Users depress the jar lid through a pad made of two 2~mm single-wall corrugated cardboard pieces. The pad showed wear only after thousands of uses and was replaced partway through the study. The controller supports one-handed interaction, works with all HMD models, and provides adequate MVR input performance. 

Pressing the button emits a click sound (see Figure \ref{fig_controller}). We filtered this with 2.5 kHz high-pass and 12 kHz low-pass filters to remove microphone audio low rumbles and high hisses. If the system detected noise that passed these filters and was louder than a minimum threshold, the audio spectral data was scanned for a spectral peak magnitude. The software recognized a sound as a button click if it had a 20--350~ms duration, peak magnitude of 0.001–0.03, an average loudness of 0.02–0.1, and the largest spectral value occurred within specific time windows early in the sound. The time windows were determined through pilot testing to ensure sounds registering as clicks matched the button click's characteristic waveform. Lastly, we employed a 300~ms debounce time between click detections to prevent false double clicks. Pilot testing found that the system is not susceptible to background "office" noise, including feet shuffling, keyboard typing, or talking. However, noises like door slams and loud speech can falsely trigger a click event.

\subsection{Steering} \label{sec: System-design-Steering}
For gaze- and controller-directed steering, travel direction followed the HMD's and controller's forward vector, respectively, with y-axis values set to zero. With both input methods, we adapted an established technique that maps controller tilt to movement speed, leveraging the advantages of motion-tracked velocity control discussed in Section \ref{sec: related-work-steering}. This button-free approach provides continuous input while avoiding MVR's hardware limitations. 

\begin{figure}[!t]
\centering
\includegraphics[width=.85\linewidth]{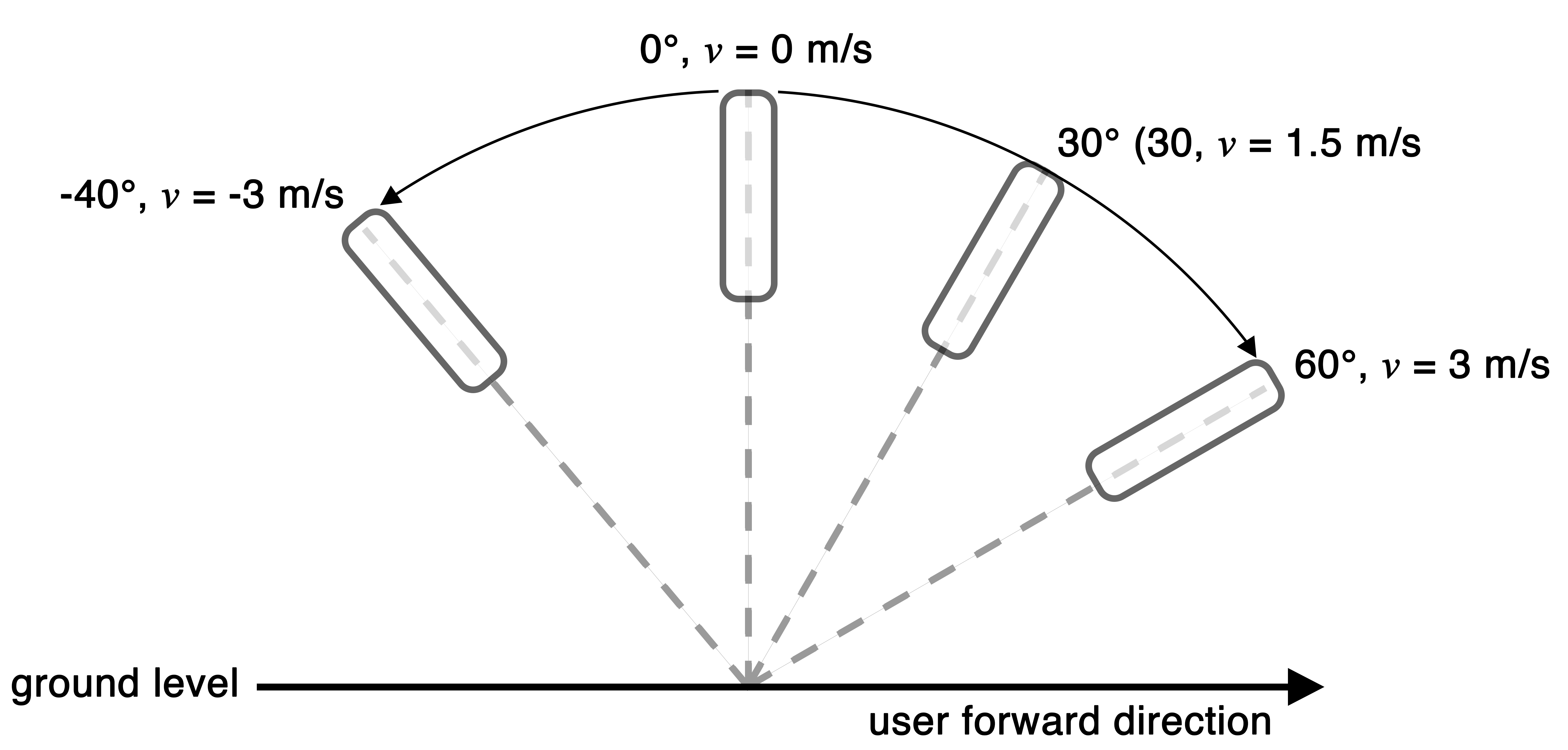}
\caption{Controller-pitch speed mapping, showing maximum forward and backward speed, and rest pose at 0°.}
\label{fig_speed-control-diagram}
\Description{Diagram showing speed values when the controller is held at -40, 0, 30, and 60 degrees relative to ground level.}
\end{figure}

The technique mapped speed to controller pitch like a gas pedal (see Figure \ref{fig_speed-control-diagram}). Pitch was measured as a signed value from the upright controller orientation (i.e., 0°). %
Holding the controller within ±10° stopped movement. Forward pitch (>0°) increased forward speed to a maximum of 3~m/s \cite{ZielaskoTestBed, Christou2017} at 60°, determined through pilot testing (Section \ref{sec: Pilot-Study}). Backward controller pitch (i.e., towards the user) increased reverse speed to the same maximum at -40° pitch; greater backward tilts were ergonomically unsuitable. %

At each frame, we calculated speed using a \textit{SmoothDamp} function, a spring-damper model that smoothly accelerates or decelerates towards the target speed, yielding more natural motion. Speeds below 0.05~m/s were clamped to 0~m/s to prevent jitter. If controller tracking was lost, speed (and direction for controller-directed steering) remained constant for 1.5~s before slowing to a stop. For backward movement, this slowing began after 200~ms since backward movement was harder to control and more likely unintended.

\subsection{Teleportation}
Teleportation used a projectile arc originating at the controller tip for controller-directed teleportation or the camera (HMD) position for gaze-directed teleportation. Based on prior work \cite{Rupp2026}, the projectile's initial velocity ($\vec{v_0}$) was 14 m/s along the aiming direction. Its positions ($\vec{p}(t)$) were sampled at $f=30$ points with timestep $\Delta t=0.1$~s via the kinematic equation $\vec{p}(t)= \vec{p}_0+ \vec{v_0}(t)+\frac{1}{2} \vec{g}t^2$ where $\vec{g}=(0,-9.8,0)$ m/s$^2$. This displayed an arc that ended at its first collision with the environment. Ground collisions yielded valid teleport positions marked with a reticle (Figure \ref{fig:conditions}b), whereas collisions with walls or decorative objects were invalid. Pressing the button at a valid teleport position instantly teleported the user without visual transition or audio feedback, since the button provided a real sound. For gaze-directed teleportation, since the origin was the user's head, the first 10\% of the arc faded from transparent to opaque white.

Teleport distance peaked at approximately 20~m when the tracked source (HMD or controller) was aimed 45° above horizontal. Teleport range decreased at shallower (<45°) or steeper (>45°) angles, reaching zero at 90° (held straight up). We added a 300~ms cooldown after each teleport to prevent rapid reactivation, based on prior work \cite{Narbayev2025} and our own pilot testing, described below.

\subsection{Pilot Study} \label{sec: Pilot-Study}
To select a speed control method for the main study and perform a preliminary assessment of our proposed technique, we conducted a pilot study comparing it (Section \ref{sec: System-design-Steering}) with automated speed control. We recruited eight participants, including researchers from our lab and strangers from other departments. The experiment was approved by our university ethics board.

We used a 2 × 2 within-subjects design with independent variables Control Type (Gaze, Controller) and Speed Control (Automated, Pitch), yielding four conditions. Participants completed a shortened version of the main study task (Section \ref{sec: methodology-task-description}) on a virtual path about one-quarter the main study path length. Temple buildings lined both sides of the path, with 8~m gaps between them (Figure \ref{fig:conditions}). %
Participants used steering to travel to the path's end while collecting five coins distributed among the temples. All conditions used the same path with different coin locations.

With automated speed, velocity was 3 m/s, decreasing to 2.5~m/s within 0.75~m of a wall. Pitch-control operated as described in Section \ref{sec: System-design-Steering}, except that controller tracking loss caused motion to continue in the last known direction, consistent with standard MVR steering implementations.

Upon completing all conditions, participants provided feedback on the speed control options. Six of eight preferred Pitch across control types. One felt Pitch paired well with controller-directed steering because the coupled movements felt natural and helped maintain controller tracking. Two described the pitch-speed mapping as intuitive, while six valued greater speed control, reporting that automated speed made avoiding walls and stopping for coins harder. One participant felt Pitch reduced cybersickness, since they could slow down when feeling symptoms. Pitch yielded fewer mean wall collisions (2.5 vs. 3.1), but lower mean speed (1.65 vs. 2.98 m/s); coin collection rate was comparable at 48\% (automated) or 50\% (speed-control). Based on these results, we selected Pitch for the main study.

\begin{figure}[t]
\centering
\includegraphics[width=.85\linewidth]{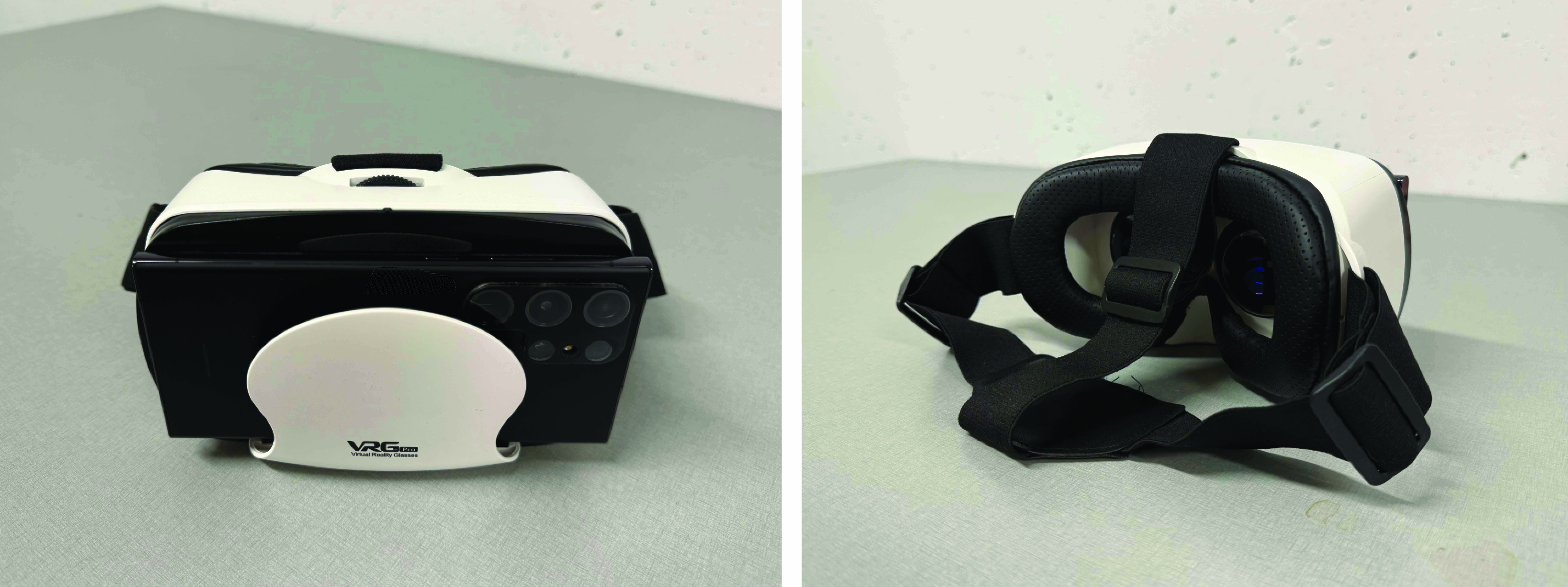}
\caption{VRG Pro HMD with smartphone inserted.}
\label{fig_hmd}
\Description{Plastic headset with head strap. An uncovered smartphone is slotted into its front.}
\end{figure}
\section{Methodology}
We conducted an experiment based on an open-source testbed \cite{ZielaskoTestBed}. %
The experiment was approved by our university ethics board.

\subsection{Participants}
We recruited 28 participants (17M, 11F) aged 18–35 (\(M = 21.9, SD = 4.67\)). All were right-handed, had normal or corrected-to-normal vision, and had previously used 3D-tracked pointing devices (e.g., VR controllers or Wii remotes). Three had no prior VR experience and one had only used cardboard VR, while 52\% self-reported as novice or expert VR users.

\subsection{Apparatus}
We used a Samsung Galaxy S23 Ultra running Android OS 14, with a 6.8”, 1440 × 3088 px ($\sim$500 PPI) 120~Hz display and a 200~MP rear camera with a 24~mm focal length. The smartphone's %
internal sensors provided 3DOF head orientation for viewpoint control. The phone was inserted into a VRG Pro (Figure \ref{fig_hmd}), an MVR HMD with a $\sim$80° horizontal FOV and IPD adjuster to ensure clear stereo viewing. The cardboard controller (Figure \ref{fig_controller}) served as the pointing device.
\begin{figure}[t]
    \includegraphics[width=\linewidth]{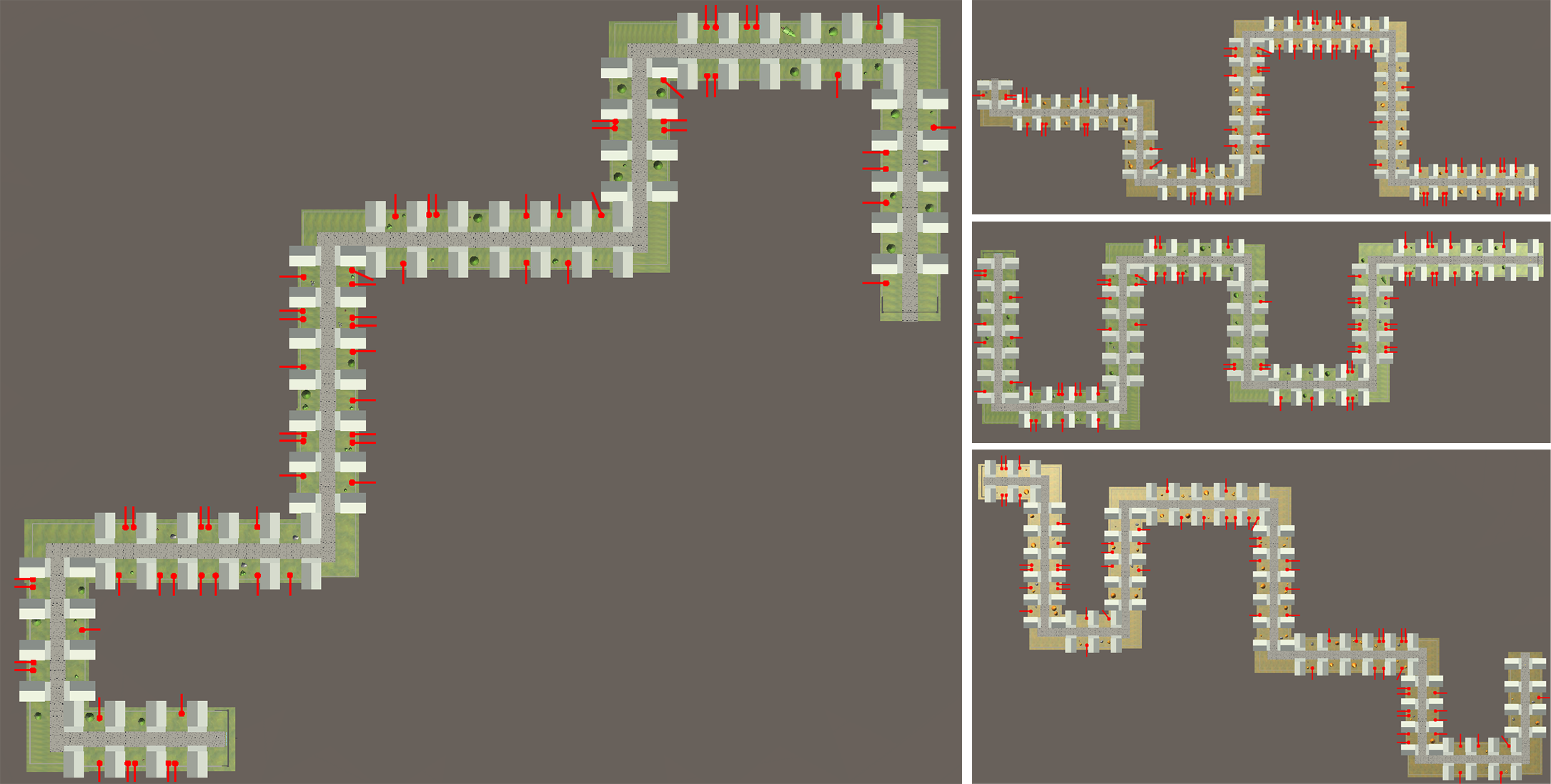}  
    \caption{The four paths used in the experiment. Red arrows indicate coin placements.}
    \label{fig_maze-paths}
    \Description{Birds-eye view of pathways. Paths have one possible route and are made of multiple left and right turns, making 90-degree angles. Coins are equally dispersed along paths and in between buildings in groups of one or two.}
\end{figure}

\subsubsection{Virtual Environment}
\begin{figure}[!t]
\centering
    \includegraphics[width=\linewidth]{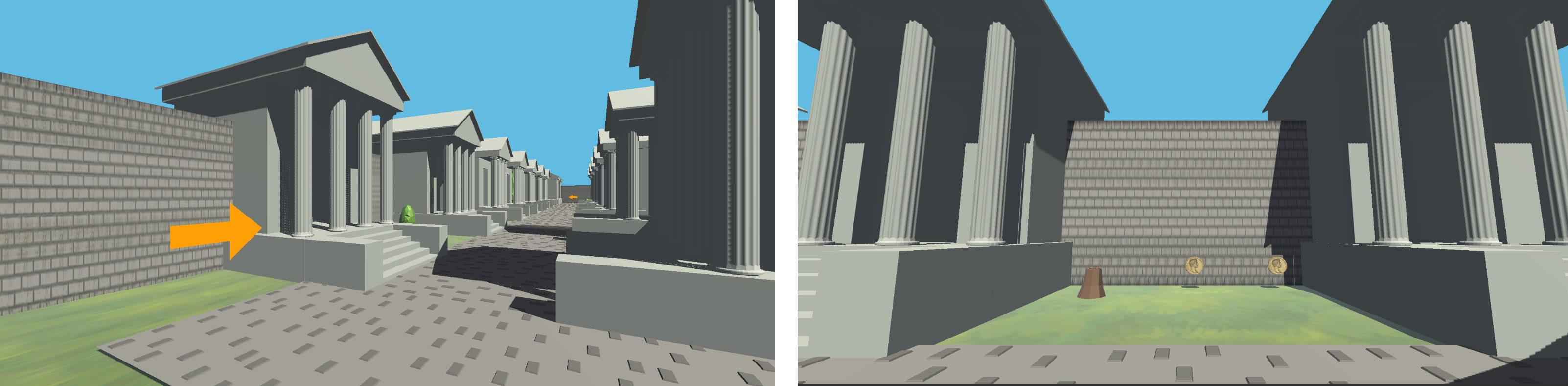}
    \caption{Task environment showing arrows as wayfinding cues (left) and coins (right).}
    \label{fig_environment}
    \Description{(Left) VR user view of a corner of a pathway. A large orange floating arrow points user in correct direction. (Right) Two floating coins in between two buildings on side of path.}
\end{figure}

We developed the software and VE in Unity 6000.2.8f1 using Vuforia (v.11.4.4) for camera tracking and Google XR (v.1.30.0) for stereo rendering. We adapted four of six unique pathways from a previous testbed \cite{ZielaskoTestBed}, assigning a different path layout to each condition (Figure~\ref{fig_maze-paths}). %
The paths were 5~m wide, single-directional, and approximately 735--780~m. %
Lining the paths were inward-facing, non-enterable temples, with 8~m gaps between them, forming 7~m-deep alleyways; both temples and alley walls supported collisions. %
The path turned 90° left or right after every 3–7 temple pairs and ended at a floating, chest-height yellow finish line. Each pathway had 64 alleyways and a total of 67 coins, with 0--2 coins floating slightly above the ground in each alleyway (Figure~\ref{fig_environment}). Participants collected coins by overlapping a gaze-controlled ring cursor with them for 50~ms, which triggered a sound effect upon collection.

To improve visuals and help participants maintain spatial context after teleportation \cite{Bowman1997}, we added grass and natural elements beside the pathways, and randomized the building colours among three neutral tones. %
Arrow cues at each turn supported wayfinding (Figure \ref{fig_environment}).

\begin{figure}[t]
    \includegraphics[width=\linewidth]{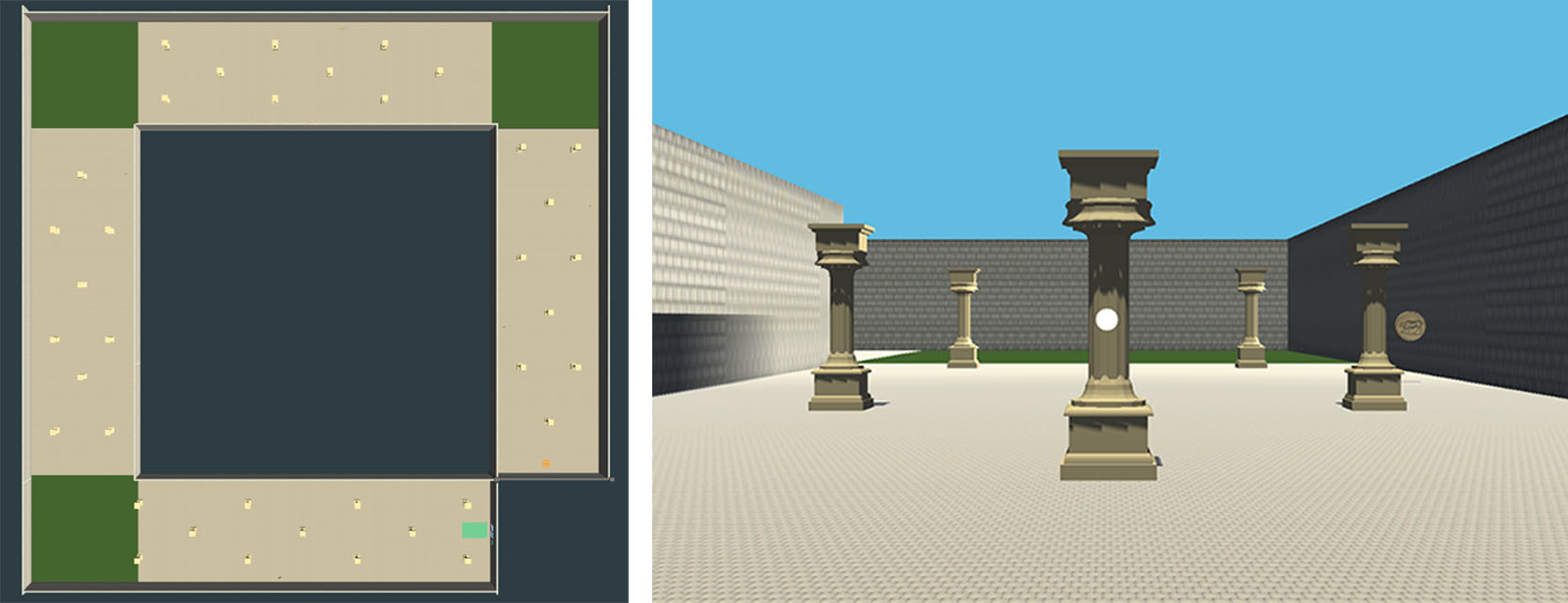}  
    \caption{Training area, Bird's-eye and first-person views.}
    \label{fig_training-area}
    \Description{Training area pathway forms a square. Pillars are evenly spaced on pathway as obstacles. Each corner of the square path is a green zone without pillar obstacles.}
\end{figure}

We used a separate training VE consisting of a square path with column obstacles and coins (see Figure \ref{fig_training-area}). Participants practiced each travel technique by weaving through the columns, with the active travel technique changing each turn.

\subsubsection{Task} \label{sec: methodology-task-description}
Participants began at one end of the pathway and received written instructions repeating the researcher's verbal instructions. Pressing a start button initiated a 3-second countdown, after which locomotion was enabled. Data recording began when participants left the starting position with steering techniques, when the countdown ended during gaze-directed teleportation, or when the controller was first detected during controller-directed teleportation.

We recorded collisions with nature elements, temples, and exterior walls. %
Continuous contact counted as just one collision. %
Participant path sampling technique differed between the two travel techniques: steering positions were sampled every 5~s, while every teleport endpoint was logged. We used different sampling techniques between teleportation and steering to reflect the different temporal structures of the techniques: steering is continuous, while teleportation is discrete. Smartphone logging load was also a concern. %

\subsection{Design}
We employed a 2 × 2 within-subjects factorial design with independent variables Control Type (Gaze-directed, Controller-directed) and Travel Technique (Steering, Teleport). The four resulting conditions were Gaze-Teleport (\textbf{GT}), Gaze-Steering (\textbf{GS}), Controller-Teleport (\textbf{CT}), and Controller-Steering (\textbf{CS}). We counterbalanced condition order using a balanced Latin square. %
The sequence of the four environments was kept constant across all participants.

Dependent variables included travel time (s), wall collisions (count), effective speed (m/s), coins collected (count), and teleport count, each recorded every 25~m of travel and upon crossing the finish line. This 25~m interval provided approximately 30 samples per path, balancing repeated measurement with meaningful progression. Each dependent variable was recorded an average of 24.6 times during teleportation and 33.7 times during steering. We also calculated the path efficiency index (PEI) once per path as the ratio of the actual path length to the optimal rectilinear path (ORP) length. Subjective measures included the short version of the User Experience Questionnaire (UEQ-S) \cite{UEQ-S-2017}, an end questionnaire, and the Fast Motion Sickness (FMS) Scale. We selected the FMS over the Simulator Sickness Questionnaire (SSQ) \cite{SSQ} primarily due to time constraints; the FMS is an established VR cybersickness measurement tool \cite{Zielasko2026} that correlates strongly with the SSQ scores \cite{FMS2011}.

\begin{figure*}[t]
    \centering
    \begin{minipage}[t]{0.33\textwidth}
        \centering
        \includegraphics[width=\textwidth]{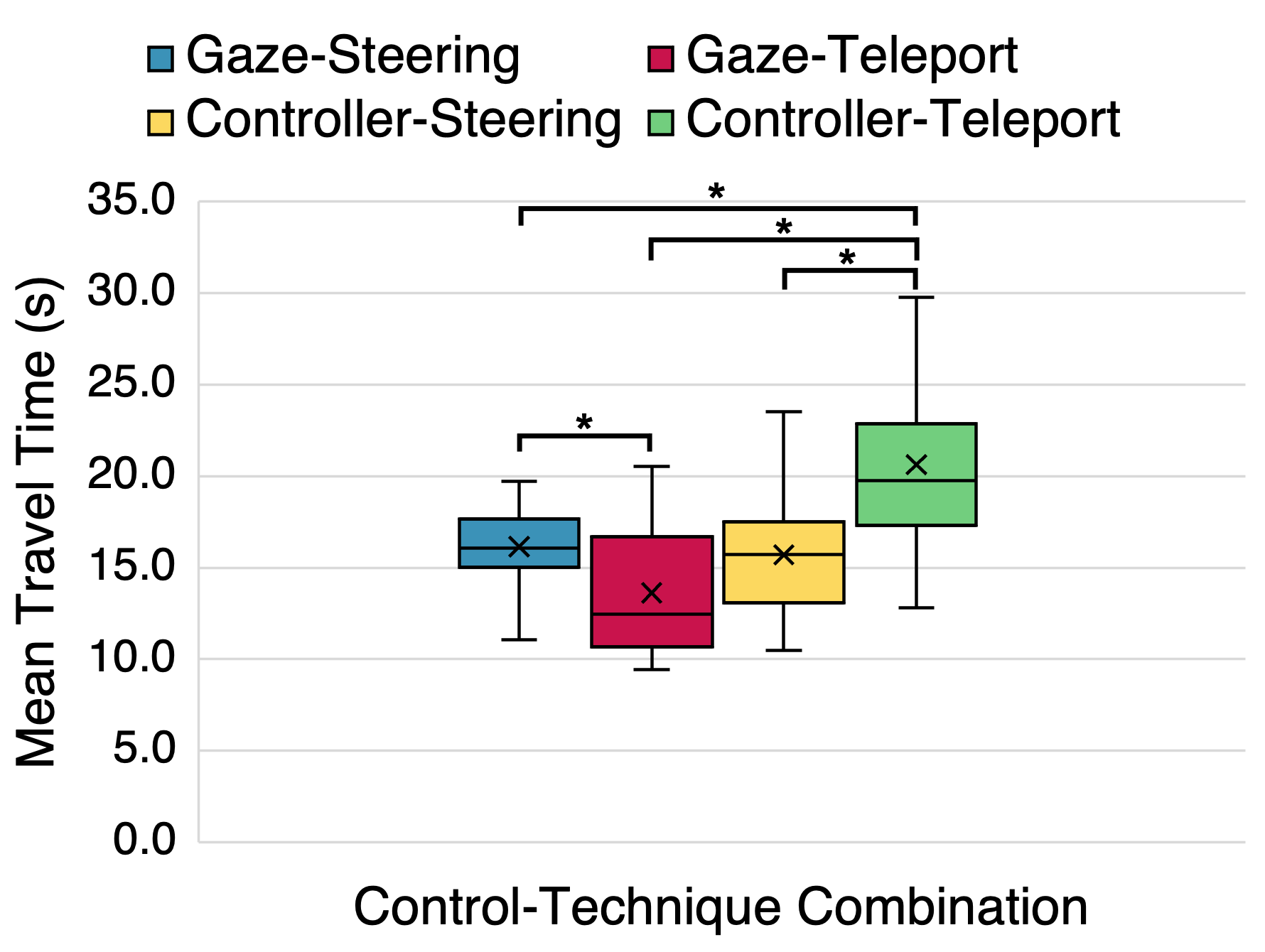}
        \caption{Mean travel time by condition. Brackets indicate \textit{p} \(<\) .05.}
        \label{fig_Travel-Time-Chart}
        \Description{Box and Whisker plot showing Controller-Teleport had significantly slower mean travel times than all other conditions. Gaze-Teleport was the fastest technique and significantly faster than Gaze-Steering.}
    \end{minipage}\hfill
    \begin{minipage}[t]{0.32\textwidth}
        \centering
        \includegraphics[width=\textwidth]{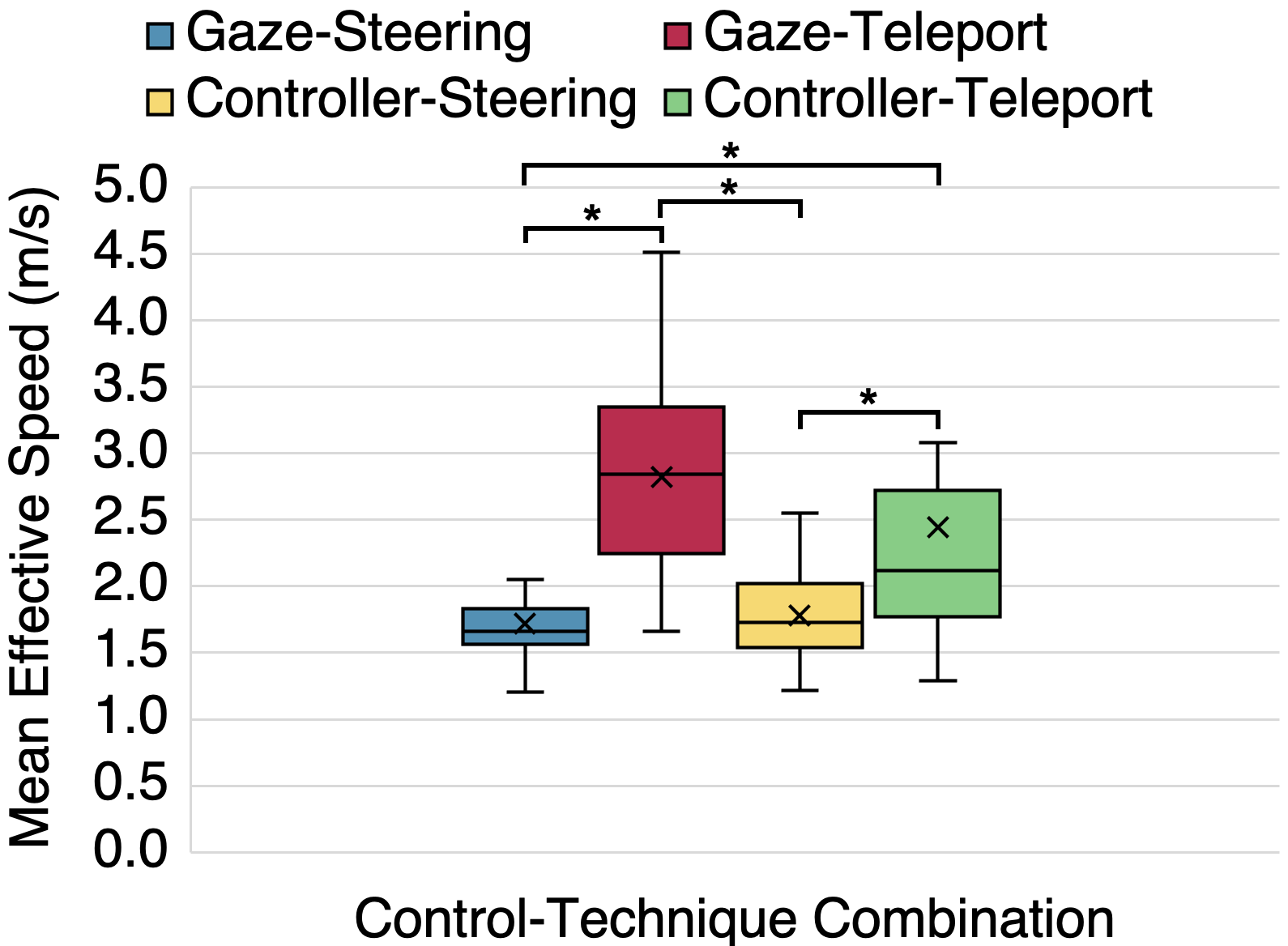}
        \caption{Effective travel speed by condition. Brackets indicate \textit{p} \(<\) .05.}
        \label{fig_Speed-Chart}
        \Description{Box and Whisker plot showing Gaze-Teleport had the fastest mean travel times, followed by Controller-Teleport, Controller-Steering, than Gaze-Steering.}
    \end{minipage}\hfill
    \begin{minipage}[t]{0.3\textwidth}
        \centering
        \includegraphics[width=\textwidth]{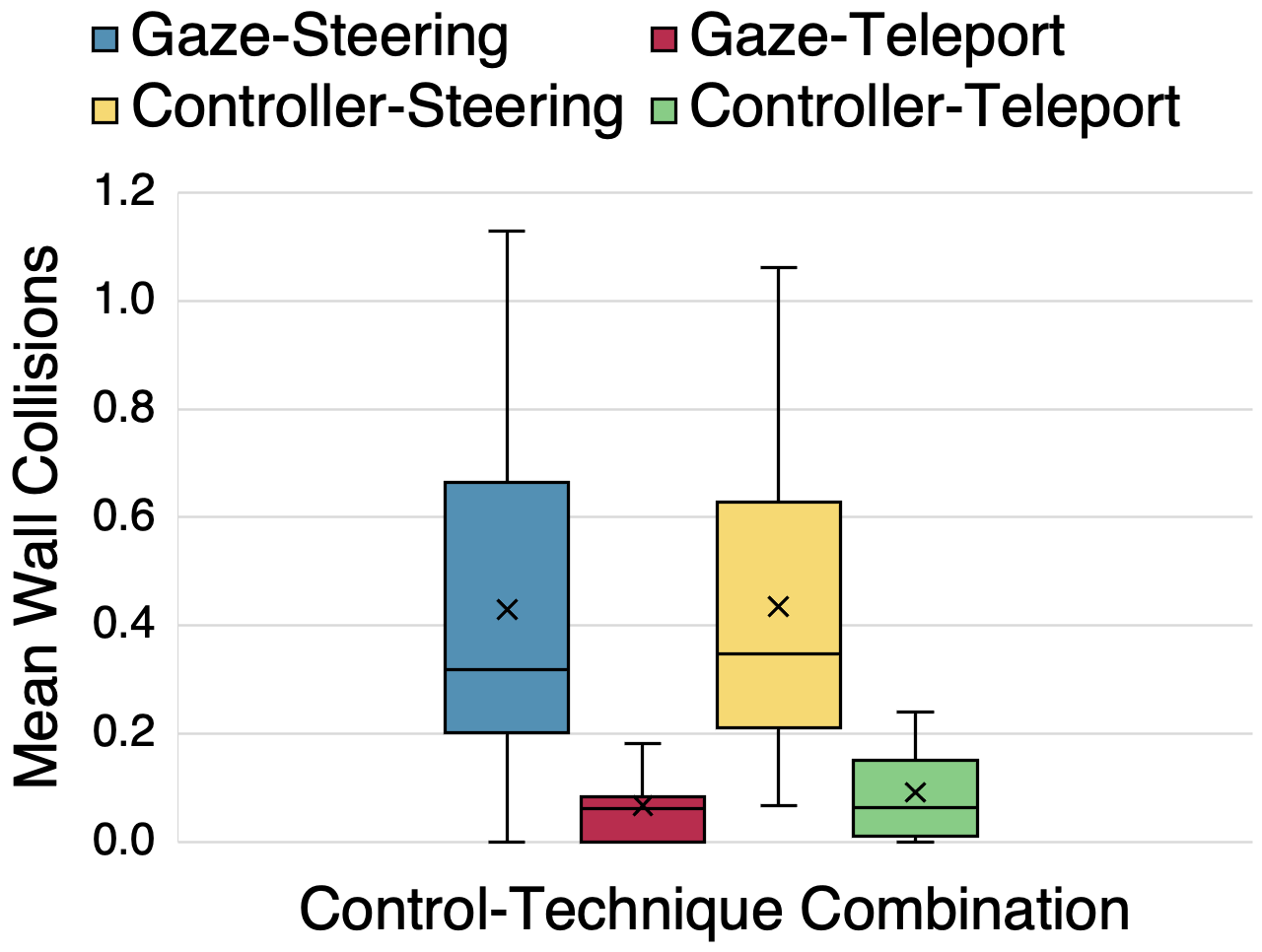}
        \caption{Mean wall collisions by condition.}
        \label{fig_Collisions-Chart}
        \Description{Box and Whisker plot showing steering techniques yielded similar average wall collision counts, both being noticeably, but not significantly, higher than the teleport techniques.}
    \end{minipage}
\end{figure*}
\subsection{Procedure} \label{sec: method-procedure}
The experiment lasted 60 minutes and began with informed consent and demographic questions, followed by an explanation of the task, conditions, and controller. Participants wore the HMD while seated in a swivel chair in a well-lit lab. %
Participants then practiced each travel technique by traversing a training path. Training was self-paced and ended when participants felt confident with all techniques. For the main task, we instructed them to cross the finish line as quickly as possible while avoiding collisions with walls or obstacles, collecting all the coins, and staying near the path centre. %

We also explained the FMS procedure; every minute, we verbally asked participants to "rate your current level of sickness on a scale from 0 to 20", where 0 indicated no sickness and 20 extreme sickness. Following the FMS Scale questionnaire \cite{FMS2011}, participants were instructed to focus on nausea and stomach discomfort, ignoring boredom and fatigue. In teleport conditions, we told participants to remain quiet unless necessary and to answer FMS questions softly to avoid accidental button triggering. After each condition, participants removed the HMD to complete the UEQ-S, providing a break averaging 1~min and 39~s. Mean time spent in VR ranged from 5~min 45~s to 9~min across conditions. During breaks, we also asked the following presence and perceived physical demand questions, following prior procedures \cite{ZielaskoTestBed}:
\begin{itemize}
\item \emph{How present ('really there') did you feel during the last level?}– [Not at all (1), . . . , Absolutely (4)]
\item \emph{Was the travel method physically demanding to use?}– [Not at all (1), . . . , Absolutely (4)]
\end{itemize}
After completing all conditions, participants filled out an overall preference questionnaire, where they ranked their agreement with the following statements:
\begin{itemize}
    \item \emph{Button registration accuracy (unregistered clicks) notably impeded my ability to complete the task}– [Not at all (1), . . . , Completely (5)]
    \item \emph{The controller tracking quality notably impeded my ability to complete the task}– [Not at all (1), . . . , Completely (5)]
\end{itemize}
The questionnaire also asked participants to rank each travel technique and provide written comments on what they liked or did not like about using their head or the controller to operate travel. 
Upon finishing the experiment, we compensated participants with \$15 CAD and sanitized the HMD ahead of the next participant.

\subsection{Hypotheses}
Based on prior work, we formed the following hypotheses:

\textbf{H1}: \textit{Teleportation will yield shorter travel time, fewer wall collisions, and better path efficiency} \cite{Buttussi2019, Dylong2026, Clifton2020, Christou2017}.

\textbf{H2}: \textit{Gaze-directed teleportation will yield shorter travel time and be preferred over controller-directed teleportation.} The button previously performed well with gaze-based selection, outperforming a controller \cite{Grinyer-GI26}. %
We expected head-gaze pointing and button activation to support efficient teleportation, whereas controller pointing would require slower, more precise movements with teleportations closer to the intended destination.

\textbf{H3}: \textit{Controller-directed steering will be preferred over gaze-directed steering.} Consistent with prior work \cite{Hegde2016, Grinyer2023, Grinyer-TVCG}, participants will find steering easier with the controller since this keeps the controller in the camera FOV and better integrates with pitch-based speed control.

\textbf{H4}: \textit{Participants will collect more coins while steering} \cite{Christou2017}.

\textbf{H5}: \textit{Controller-directed steering will produce the highest presence while gaze-directed steering will be most physically demanding} \cite{Clifton2020}.

\begin{figure*}[t]
    \centering
    \begin{minipage}[t]{0.27\textwidth}
        \centering
        \includegraphics[width=.99\linewidth]{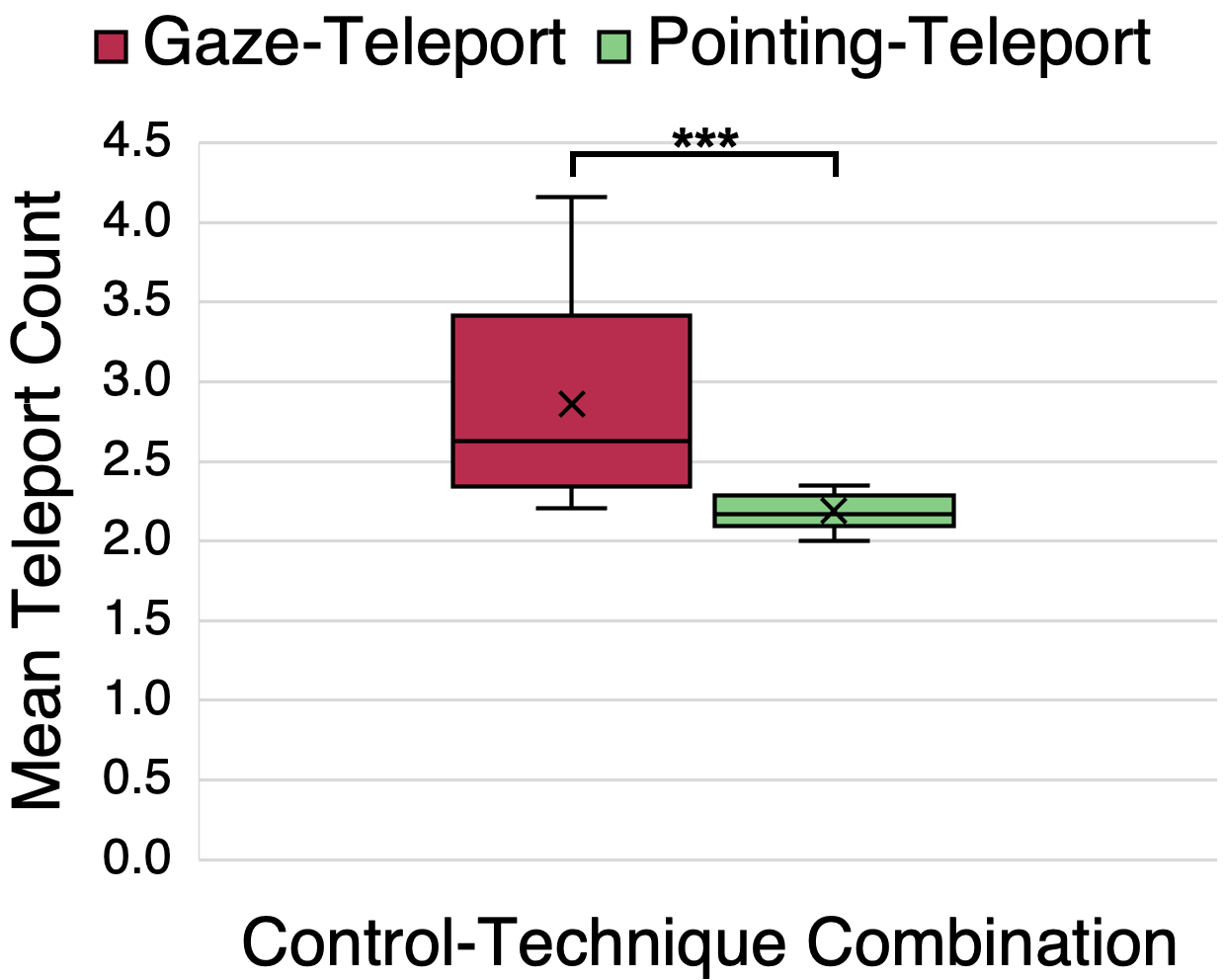}
        \caption{Mean teleport count by condition. Bracket indicates \textit{p} \(<\) .001.}
        \label{fig_TeleportCount}
        \Description{Box and Whisker plot showing Gaze-Teleport yielded a significantly higher teleport count on average, along with a less consistent number of teleports when compared to Pointing-Teleport.} 
    \end{minipage}\hfill
    \begin{minipage}[t]{0.34\textwidth}
        \centering
        \includegraphics[width=.95\linewidth]{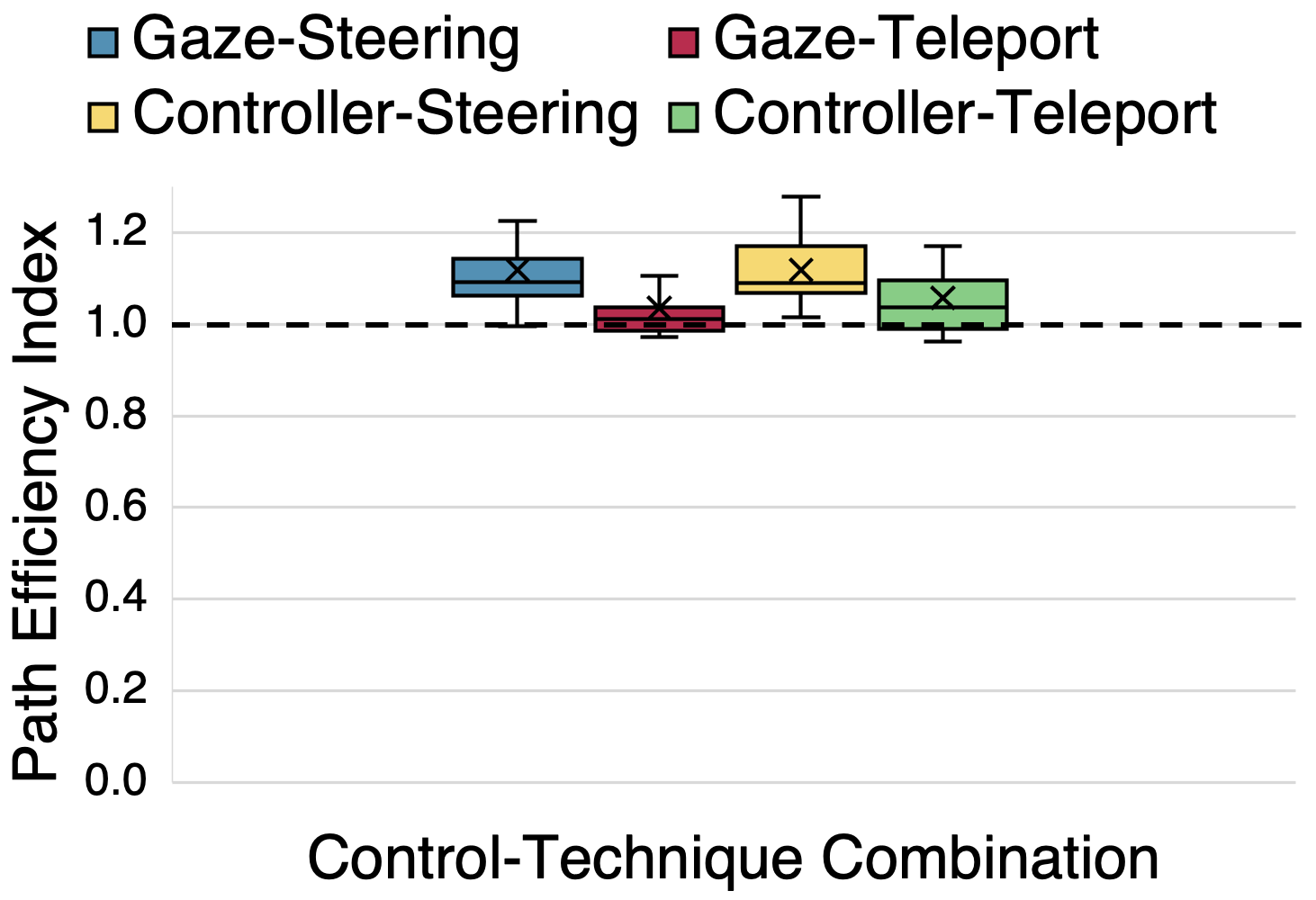}
        \caption{Path Efficiency Index by condition. Dotted line is ORP baseline.}%
        \label{fig_PEI-Chart}
        \Description{Box and Whisker plot showing all conditions had comparable path efficiency indexes (PEI). Gaze-Teleport has the lowest and closest path efficiency index to 1.0.}  
    \end{minipage}\hfill
    \begin{minipage}[t]{0.35\textwidth}
        \centering
        \includegraphics[width=\textwidth]{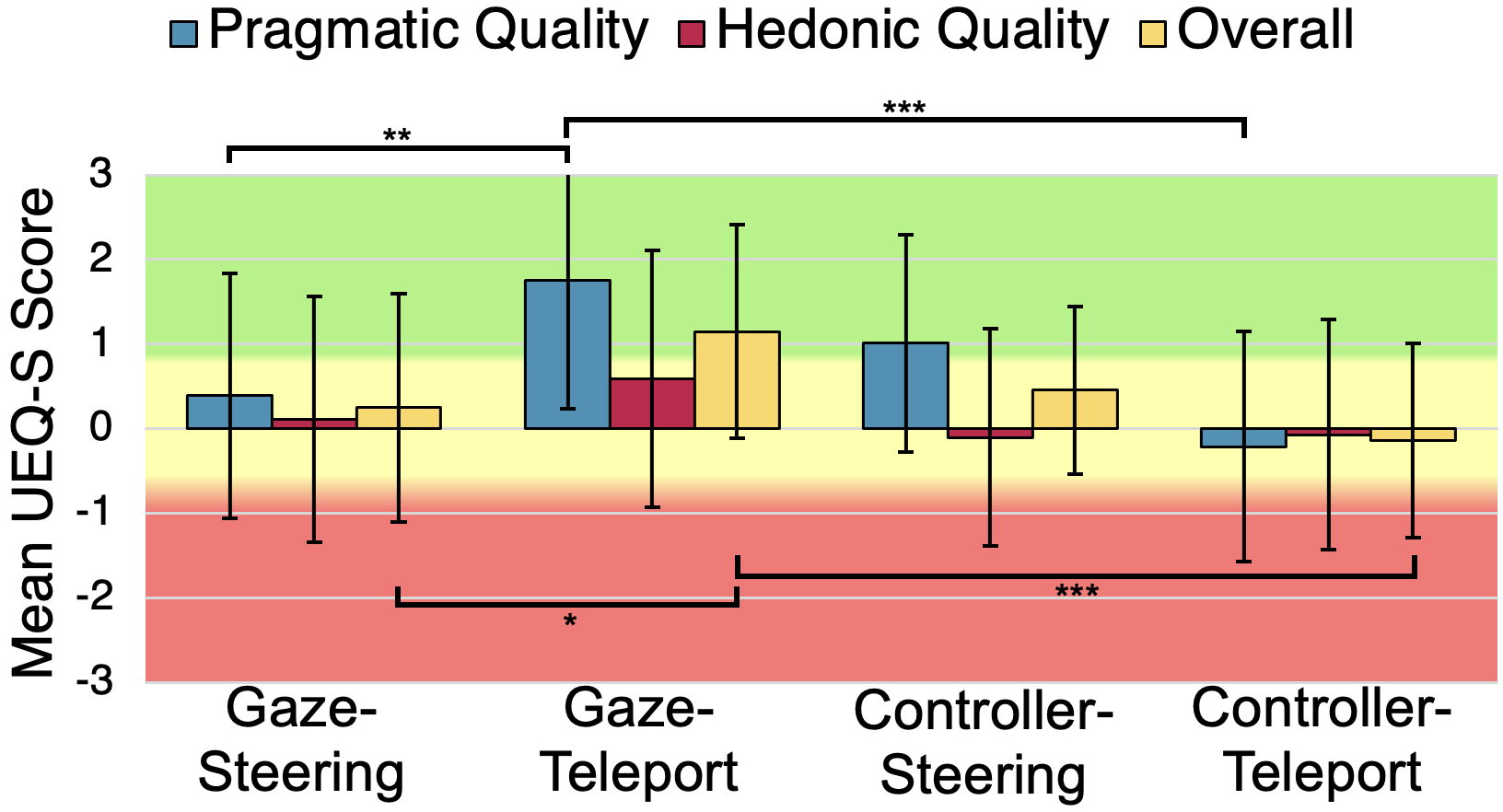}
        \caption{Mean UEQ-S scores. Colour bands show UEQ-S quality benchmarks. Brackets indicate \textit{p} \(<\) .01 (**) and \textit{p} \(<\) .001 (***).}
        \label{fig_UEQ}
        \Description{Clustered bar chart showing Gaze-Teleport has a notably higher pragmatic quality score, significantly higher than Controller-Teleport, and higher scores overall. Gaze -Steering and Controller-Teleport have scores slightly above or below zero, respectively.}
    \end{minipage}
\end{figure*}
\section{Results}
\subsection{Performance Metrics}
We used participant averages of all data points recorded every 25 meters of travel. We performed Lilliefors normality tests ($\alpha$ = .05) for each metric. We analyzed metrics that were not normally distributed using two-way ART RM ANOVA, and normally distributed metrics with two-way RM ANOVA. Post-hoc analyses used Tukey HSD. There was no significant effect of condition order on completion time (\(F_{3,24}\) = .682, \(p >\) .5, \(\eta_{p}^{2}\) = .08, \(pow\) = .04), suggesting counterbalancing was effective.

\subsubsection{Travel Time}
Two-way RM ANOVA revealed a significant main effect of Control Type on travel time (\(F_{1,27}\) = 36.82, \(p <\) .0001, \(\eta_{p}^{2}\) = .58, \(pow\) = 1.0); on average, participants were faster using Gaze controls (15.21~s) over the controller (17.93~s). A significant interaction effect (\(F_{1,27}\) = 46.42, \(p <\) .0001, \(\eta_{p}^{2}\) = .63, \(pow\) = .98) and post hoc testing revealed CT had higher travel times than all other conditions (see Figure \ref{fig_Travel-Time-Chart}), supporting \textbf{H2} and partially rejecting \textbf{H1}.

\subsubsection{Effective Speed}
Effective speed is the straight-line distance travelled over time. The maximum travel speed with steering conditions was 3~m/s, while the maximum teleport distance was 20 m. Two-way RM ANOVA found a significant main effect of Travel Technique on effective speed (\(F_{1,27}\) = 31.17, \(p <\) .0001, \(\eta_{p}^{2}\) = .54, \(pow\) = .99). On average, participants moved 0.86~m/s faster under Teleport conditions. A significant interaction effect (\(F_{1,27}\) = 5.43, \(p <\) .05, \(\eta_{p}^{2}\) = .17, \(pow\) = .97) and post hoc tests revealed GS and CS were significantly slower than both teleport conditions (see Figure \ref{fig_Speed-Chart}).

\begin{figure*}
\centering
\includegraphics[width=\linewidth]{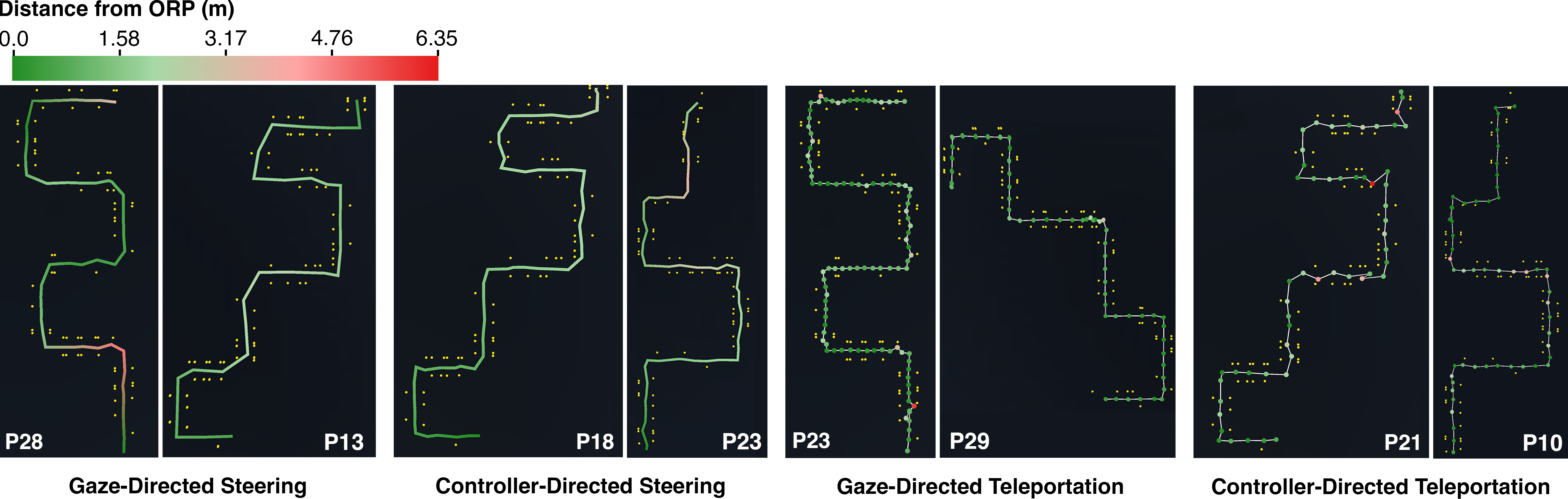}
\caption{Example participant paths for each condition. Steering paths are lines and use every other logged data point for visual clarity; teleport locations are dots, colour indicates distance from ORP, and yellow dots are coin locations.}
\label{fig_user-paths}
\Description{Example paths coloured using a spectrum of green to red. Darker green indicates closeness to the optimal path, while red indicates a farther distance. A Gaze-Steering path has the most red, while Gaze-Teleport has the most green. The other conditions are mainly comprised of dark green, light green, and a neutral in-between colour.}
\end{figure*}
\subsubsection{Wall Collisions}
On average, less than one wall collision occurred per 25~m of travel regardless of condition. See Figure \ref{fig_Collisions-Chart}. Two-way ANOVA found a significant main effect of Travel Technique on wall collisions (\(F_{1,27}\) = 43.17, \(p <\) .0001, \(\eta_{p}^{2}\) = .62, \(pow\) = .97). Collisions occurred far more frequently with steering (0.44 per 25 m) than teleportation (.08 per 25 m), supporting \textbf{H1}.%

\subsubsection{Coins Collected}
Each path contained approximately 2.1–2.3 coins per 25~m of the ORP. Two-way ART ANOVA revealed a significant main effect of Travel Technique on coins collected (\(F_{1,81}\) = 84.54, \(p <\) .001, \(\eta_{p}^{2} =\)~.51, \(pow\) = .98); participants collected fewer coins in Steering conditions, contrary to \textbf{H4}. %
Overall, participants collected 79–89\% of 67 coins. Collapsing across Travel Technique, gaze yielded 9.55 more coins than the controller.

\subsubsection{Teleport Count}
One-way ANOVA revealed a significant effect of Control Type on teleport count (\(F_{1,27}\) = 35.24, \(p <\) .0001, \(\eta_{p}^{2}\) = .57, \(pow\) = .96). Participants teleported more often with GT ($M = 2.88, SD = .6$) than CT ($M = 2.2, SD = .13$). See Figure \ref{fig_TeleportCount}.

\subsubsection{Path Efficiency}
PEI was the ratio of the travelled path length to the ORP length, based on 9102 total recorded positions. PEI scores above 1 indicate a longer route than the ORP, whereas scores below 1 indicate a more efficient route by cutting corners at turns. %
Mean PEIs are seen in Figure \ref{fig_PEI-Chart}. Two-way ART RM ANOVA revealed a significant main effect of Travel Technique on PEI (\(F_{1,81}\) = 45.01, \(p <\) .0001, \(\eta_{p}^{2}\) = .36, \(pow\) = 1.0). Steering produced less efficient paths than teleportation, supporting \textbf{H1}. This is not surprising, as steering may cause deviation toward coins, whereas teleportation lets participants remain in place more easily. Figure \ref{fig_user-paths} shows archetypal participant paths and deviation at coins. The displayed paths correspond to the participants whose PEI scores were closest to the mean for each condition. %

\begin{figure*}[!t]
    \centering
    \begin{minipage}[t]{0.4\textwidth}
        \centering
        \includegraphics[width=\textwidth]{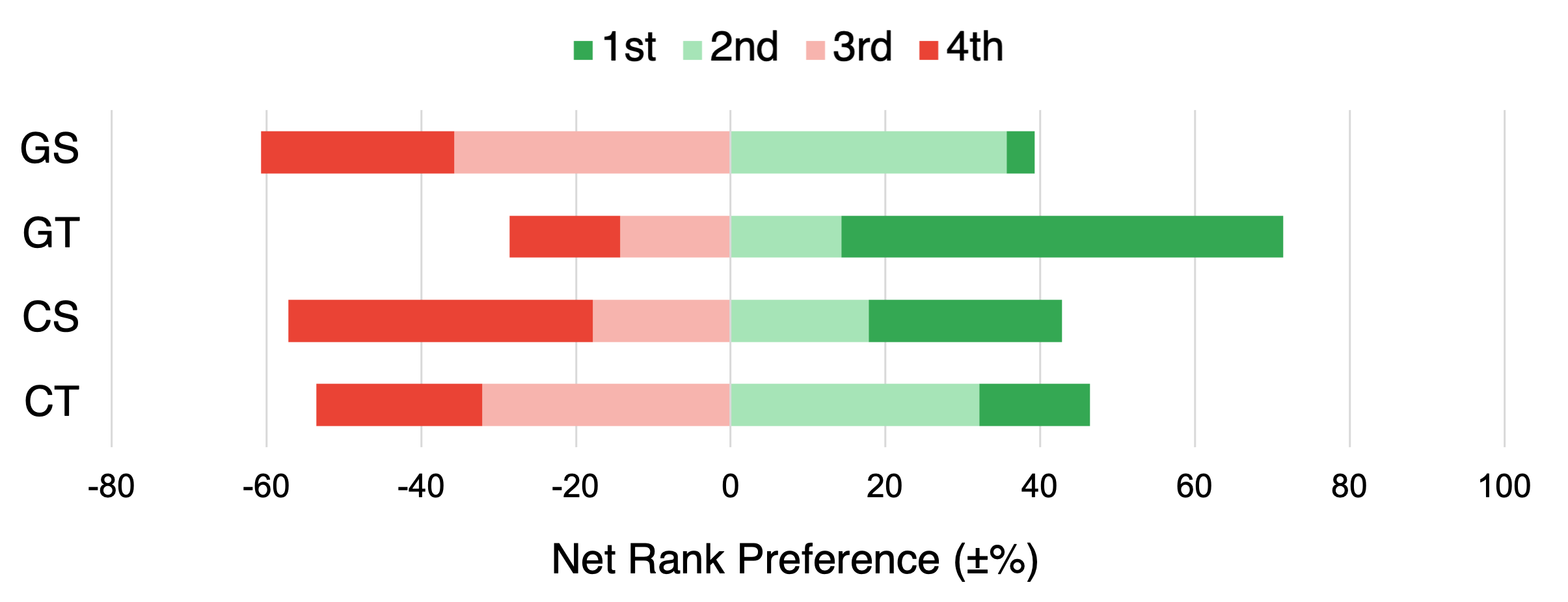}
        \caption{Participant rankings of conditions. Percentage of participants who chose each ranking.}
        \label{fig_rankings}
        \Description{Stacked bar chart showing most participants ranked Gaze-Teleport as their first choice, then Gaze-steering. Controller-Teleport had the second highest average rating while Gaze-Steering had the lowest average rating.}
    \end{minipage}
    \hspace{0.03\textwidth}
    \begin{minipage}[t]{0.56\textwidth}
        \centering
        \includegraphics[width=.49\textwidth]{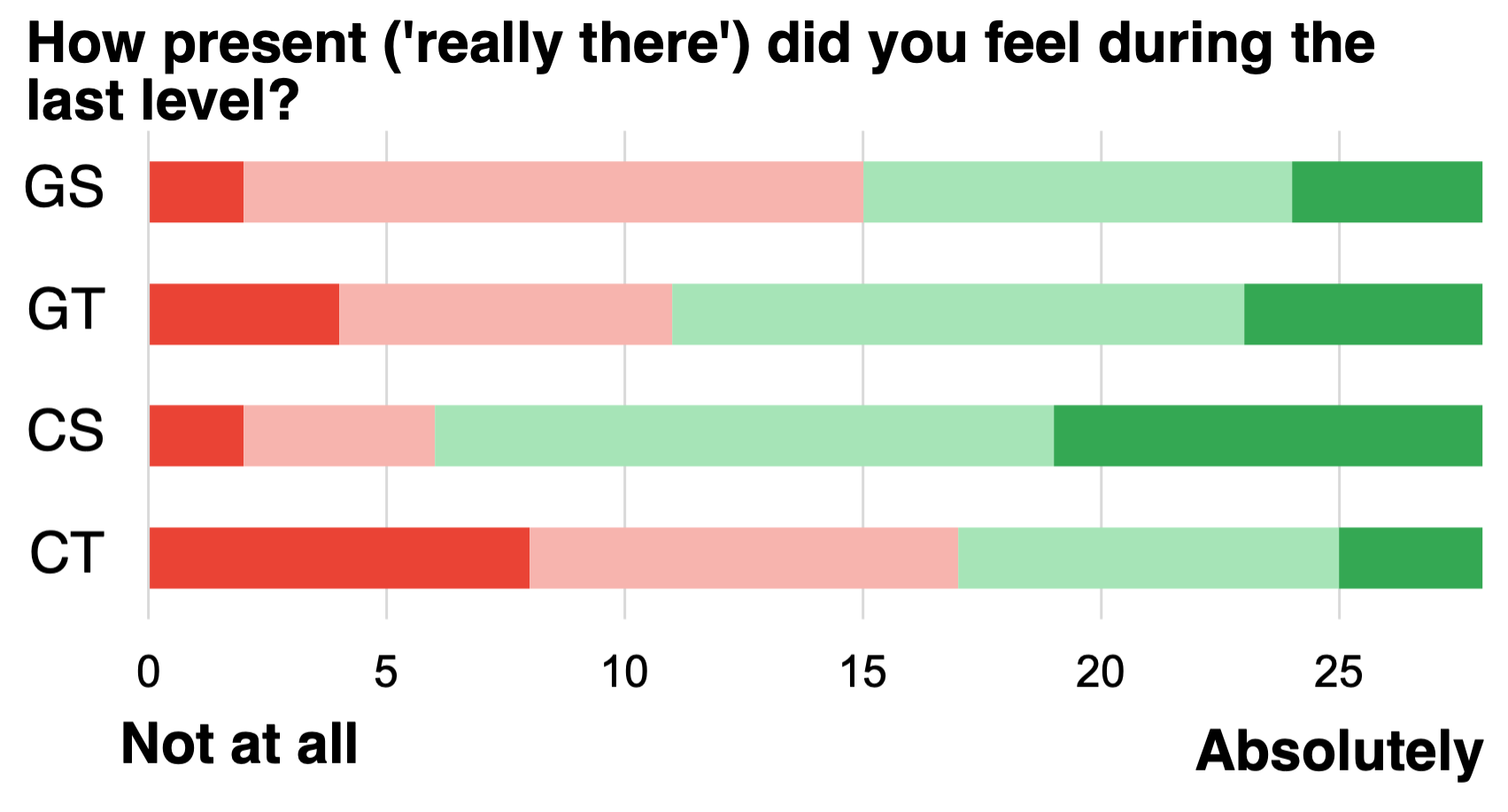}
        \includegraphics[width=.49\textwidth]{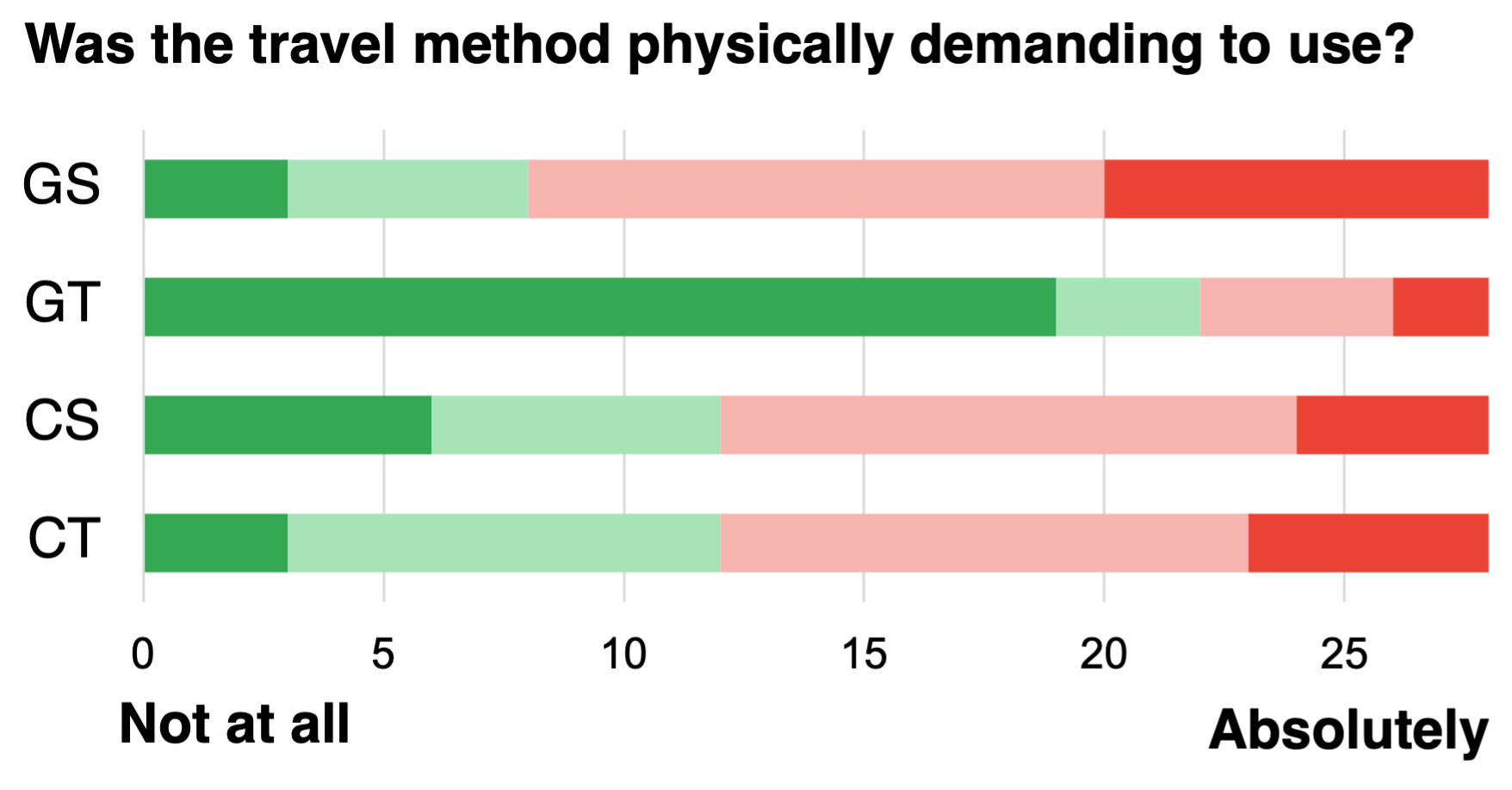}
        \caption{Participant responses to 4-point Likert scale items on presence and physical demand for all conditions.}
        \label{fig_presence-phys-demand}
        \Description{(Left) Stacked bar chart showing presence rankings on 4-point scale. The majority of participants gave Controller-Steering and Gaze-Teleport a positive rating of presence. (Right) Stacked bar chart showing perceived physical demand on 4-point scale. 22/28 participants did not perceive Gaze-Teleport as physically demanding.}
    \end{minipage}
\end{figure*}
\subsection{Subjective Feedback} \label{sec: subjective results}

\subsubsection{Fast Motion Sickness Scale}
A Friedman test on the FMS scores failed to detect a significant difference (\(\chi^2(3)\) = 0.81, \(p\) = 0.85, \(W\) = .001, \(pow\) = 0.20). We also compared within-condition cybersickness change by subtracting each participant's initial FMS score from their final FMS. This gives an approximation of cybersickness increase over time. These change scores also did not differ significantly between conditions (\(H'_{3}\) = 1.79, \(p\) = 0.62, \(W\) = .02, \(pow\) = 0.43).%

\subsubsection{Questionnaires}
As detailed in Section \ref{sec: method-procedure}, participants provided information on several questionnaires:  

\textbf{Input Device Ranking.} Participants rated how much the button and controller impeded performance on a 1–5 scale. Mean scores of 3.07 and 3.79 respectively, suggest both somewhat hindered performance, perhaps because participants were using the device for the first time.

\textbf{User Experience Questionnaire.} The UEQ-S has 8 items measuring Pragmatic Quality (usability) and Hedonic Quality (engagement, novelty) on a -3 to +3 scale. Applying the standard UEQ-S data-sanity heuristic \cite{Schrepp2023}, we excluded six highly inconsistent responses, leaving GS ($N=27$), GT ($N=28$), CS ($N=26$), and CT ($N=26$). A two-way RM ANOVA revealed a significant main effect of Control Type on overall UEQ-S score (\(F_{1,27}\) = 5.65, \(p <\) .05, \(\eta_{p}^{2}\) = .17, \(pow\) = 1.0), favouring Gaze, and a significant Control Type $\times$ Travel Technique interaction effect (\(F_{1,27}\) = 14.5, \(p <\) .001, \(\eta_{p}^{2}\) = .35, \(pow\) = 1.0). Figure \ref{fig_UEQ} shows mean scores and significant pairwise differences identified by Bonferroni-corrected paired t-tests. GT had the highest scores on all scales, significantly more so than both teleport techniques. Participant comments support these results: 11 participants described GT as “\textit{easy}”, “\textit{efficient}”, and "\textit{fast}" %
, while others described it as “\textit{practical}” (P16), “\textit{accurate}” (P8), and “\textit{intuitive}” (P28). However, P1 found it "\textit{uncomfortable}" because continuing forward required "unnatural angles" of the head.

The usability of CS was moderate (see Figure \ref{fig_UEQ}). Participants preferred decoupled movement control as it “\textit{felt natural and not restrictive}” (P1), enabled multitasking through simultaneous movement and coin searching %
(P22), and allowed a comfortable head position (P1, P24). However, participants also reported controller-based tracking instability: seven participants noted tracking issues, with three identifying the narrow FOV as a contributing factor (P1, P19, P26). Participants experienced some arm fatigue with the controller; P16 noted that it was “\textit{physically demanding to keep an arm up at all times}.”

GS and CT received mid-range UEQ-S scores, with usability (pragmatic quality) rated lowest for both conditions (Figure \ref{fig_UEQ}). Nevertheless, participants considered GS controls clear. Participant comments associated GS with discomfort and restricted movement. For example, P9 reported it caused “the most motion sickness” and %
P1 felt it “restricted” them to looking in the travel direction, resulting in unintended movement when looking for coins (P22). Comments on CT echo the tracking issues noted above. While participants valued the ability to look around, some (e.g., P1) found it difficult to reacquire the controller after looking away.

\textbf{Condition Preferences.} Participants ranked all four conditions by preference (Figure \ref{fig_rankings}) and explained their choices. A Friedman test revealed a significant difference in rankings (\(\chi^2(3)\) = 9.64, \(p <\) .05, \(W\) = .11, \(pow\) = .74). Bonferroni-corrected Conover post hoc comparisons showed that GT received significantly higher rankings, supporting \textbf{H2} and aligning with UEQ-S scores. However, \textbf{H3} was not statistically supported, although descriptive results favoured CS over GS. %
Specifically, CS was ranked first by seven participants (25\%), compared with one for GS, and appeared among the top two for 12 participants (42.86\%), compared with 11 (39.29\%) for GS. However, CS also received more last-place rankings than GS (11 vs. 7), indicating more polarized rankings. The mismatch between CT’s moderate UEQ-S score and its low preference ranking is discussed in Section \ref{sec: discussion}.

\textbf{Presence and Physical Demand.} Friedman tests revealed significant differences in presence (\(\chi^2(3)\) = 17.89, \(p <\) .001, \(W\) = 0.21, \(pow\) = 1.0) and physical demand (\(\chi^2(3)\) = 27.29, \(p <\) .0001, \(W\) = 0.33, \(pow\) = 1.0). Bonferroni-corrected Conover comparisons indicated that GT was significantly less physically demanding than all other conditions, while GS was significantly more demanding than CS. For presence, post hoc comparisons showed that CT received significantly lower ratings than GT and CS, while CS received significantly higher ratings than GS. See Figure \ref{fig_presence-phys-demand}. Together, these results were consistent with \textbf{H5} and the broader pattern observed throughout the study: gaze was better suited to teleportation, while decoupled control was better suited to steering.
 
In contrast to reports of teleportation reducing presence \cite{Clifton2020}, GT received descriptively higher presence ratings than GS, although the difference was not significant. This may be due to the large usability differences detected by the UEQ-S: the higher-performing techniques, GT and CS, also received the highest presence ratings, suggesting that usability may influence presence more strongly than travel technique.

\section{Discussion} \label{sec: discussion}
Below, we compare locomotion techniques and offer guidelines for choosing a travel technique for mobile VR.
\subsection{Teleport Control Scheme and User Strategies}
Overall, GT outperformed and was preferred to CT, supporting \textbf{H2} and consistent with past work \cite{Narbayev2025}. CT teleports were slower but more precise, whereas gaze-directed controls required less precise aiming and enabled more natural movement, as reflected by Teleport Count results and participant feedback (e.g., “\textit{[GT] made teleportation and navigating intuitive since I could just [look-and-teleport]}”, P10). The example travel paths in Figure \ref{fig_user-paths} illustrate this difference: with CT, participants frequently deviated deliberately from the ORP to collect coins. Several participants struggled with short-range teleports with CT because doing so required either pointing the controller at the ground or angling it upward. Pointing downward made aiming and previewing the destination easier, whereas angling upward was more difficult. Neither strategy was universally intuitive despite training; further practice may improve CT performance and usability. These movement pattern differences may also explain why participants collected more coins with GT, since shorter teleports supported more systematic search strategies.%

\subsection{Comparison of Teleportation and Steering}
Teleportation yielded lower path deviation (Figure \ref{fig_PEI-Chart}) and fewer wall collisions (Figure \ref{fig_Collisions-Chart}) than steering, partially supporting \textbf{H1}. %
However, CT was slower than both steering conditions, so \textbf{H1} was only partially supported. This is inconsistent with prior work \cite{Christou2017}, and may reflect the CT usability issues noted above. CT also showed a mismatch between the interval-level effective speed and cumulative travel time, likely because larger teleportation jumps inflated speed (Figure \ref{fig_user-paths}). Although CT had slower interval-level effective speed, its mean total travel time (489.3 s) was still lower than GS (540.6 s) and CS (523.4 s). This suggests that interval-based and cumulative measures produced different interpretations of performance, warranting further investigation.

Participant preferences reflected trade-offs among three themes that emerged from their comments: efficiency and speed (7 responses), reduced motion sickness (5 responses), and intuitive interaction (6 responses). Preferred techniques were “easy and efficient”, “intuitive”, or “natural”. Thus, both subjective and objective results favoured teleportation over steering, consistent with prior work \cite{Buttussi2019,Clifton2020}. 

Prior work has reported mixed results on cybersickness, with some finding steering worse \cite{Dylong2026,Christou2017}, while others found no clear difference between techniques \cite{Clifton2020,Dresel2021}. We found no significant effects for cybersickness, although some participants reported less sickness when teleporting, while others preferred steering. %
Across Control Type, 71\% identified Gaze as the main source of nausea, whereas controller use was generally associated with less discomfort. These results suggest that the Control Type may influence cybersickness more than Travel Technique in MVR, potentially helping explain conflicting findings in prior comparisons.

To contextualize our results, we descriptively compared cumulative measures to those reported by Zielasko et al. \cite{ZielaskoTestBed}, who evaluated gaze- and torso-directed steering with leaning-based velocity control using an HTC Vive Pro. Although the task and environments were closely matched, substantive differences in display, tracking, and input hardware limit this to a high-level comparison. %
Their steering conditions were faster than ours, but GT was quickest overall. Participants in our study also missed more coins while steering, whereas mean distance travelled and wall collision counts were similar across the two studies' gaze-directed steering conditions. UEQ‑S results were likewise comparable: CS slightly exceeded their gaze‑directed steering condition, while their torso‑directed steering scored highest among steering techniques. %
Overall, several metrics broadly aligned with the prior study, despite the gap in hardware fidelity. This provides preliminary evidence that low‑fidelity techniques can offer usability comparable to similar high-fidelity steering techniques. A direct controlled comparison is needed to substantiate this.

\subsection{Design Guidelines for Locomotion in MVR}
Drawing on our findings and prior work on decoupled locomotion, we propose design guidelines for primed-search travel techniques using low-fidelity immersive displays. These guidelines are scoped to our hardware setup, where controller tracking was limited by camera FOV. Christou et al. found head-decoupled teleportation suited wide-FOV displays, decoupled steering performed best in CAVEs, and gaze-directed steering was more suitable for desktop VR \cite{Christou2017, Christou2016CAVE}. Collectively, these findings suggest that the benefits of head-decoupled movement control are highly dependent on display type and FOV.

\textbf{\textit{1. Use gaze-directed control for teleportation, but not continuous steering}}. In limited-input MVR systems, coupling gaze-directed teleportation and viewpoint control allows users to aim and teleport through a coordinated action. This supports efficient and accurate teleportation with higher usability, presence, and lower physical demand. However, gaze-directed steering prevents users from looking around independently while moving. Designers should therefore favour gaze-directed control for teleportation over steering.

\textbf{\textit{2. Use tracked handheld controllers to direct view-decoupled steering}}. Although GS and CS produced comparable objective performance, CS yielded higher presence and was described as more intuitive, natural, and less restrictive. Participants also associated GS with greater motion sickness, although FMS scores did not differ significantly. Preferences for CS were polarized: some participants valued the freedom and naturalness of decoupled control, whereas tracking instability and arm fatigue led others to rank it poorly. Thus, low-fidelity handheld controllers can support steering on small-FOV displays, assuming reliable tracking can be maintained. Despite comparable FOVs, our results differed from Christou et al.'s desktop VR study \cite{Christou2017}, suggesting that display type and immersion level may influence the decoupling of movement and viewpoint control.

\textbf{\textit{3. Use tilt controls for continuous speed control}}. Participants described the pitch-based speed-control technique as intuitive and “\textit{resembling driving}” (P26), while six of eight pilot participants preferred it to automated speed control. Prior MVR studies found that users disliked automated speed and start/stop controls \cite{Powell2017,Bothen2018} while HMD buttons and joysticks provided no clearly satisfactory speed control mechanism \cite{Powell2017}. In standalone VR, leaning-based velocity control has outperformed joysticks in locomotion tasks \cite{Buttussi2019, ZielaskoTestBed, Hashemian2023, Harris2014, Marchal2011}. Together, these findings suggest that controller tilt offers a simple and easily tracked button-free orientation-based option for speed control in low-cost MVR systems.

\subsection{Potential of Lo-Fi Devices for Mobile VR}
Overall, GT's performance suggests that the audio-based button can effectively support teleportation in MVR. Controller-directed pointing adequately supported steering but performed poorly for teleportation. Nevertheless, participants valued the controller's novelty, responsiveness, light weight, and intuitive speed control, although tracking loss caused by the camera's limited FOV was a common drawback. This contrast between subjective appeal and performance suggests that low-fidelity devices warrant further exploration, particularly with improved tracking. Our prior work similarly indicates that low-fidelity prototypes using hybrid tracking systems can support effective MVR interaction \cite{Grinyer2026-Review}. %
Given MVR's potential as an accessible entry point to XR, these findings support continued development of low-cost input devices to help new users adopt VR and develop spatial UI skills.

\subsection{Limitations}
Our study has several limitations informing future work. First, the cross-study comparison with the prior testbed is approximate because the display, tracking, and input hardware differed. A direct comparison with higher-fidelity VR devices would better contextualize the observed performance differences. %
Second, the button occasionally produced false negatives, which may have affected teleportation metrics and perceived usability. %
Third, the coin-collection task introduced behavioural variability because participants differed in how they prioritized travel speed and coin collection. %
In particular, one participant attributed their discomfort to head motions required by the task. %
Finally, path trajectories were recorded differently between Travel Techniques: steering positions were sampled every 5~s, whereas every teleport endpoint was recorded. Although this procedure reflected the techniques' continuous vs. discrete temporal structures and reduced smartphone logging overhead, the unequal sampling rates may have affected trajectory-level measures and limit direct comparison across techniques.

\section{Conclusion}
Adapting Zielasko et al.'s testbed \cite{ZielaskoTestBed}, we used a low‑cost tracked controller \cite{Grinyer-TVCG} and audio‑based button input \cite{Grinyer-GI26} to compare free‑range teleportation and speed‑controlled steering with gaze- and controller-directed control. Performance and preference depended strongly on how Travel Technique and Control Type were coupled: gaze‑directed teleportation produced the best overall results, whereas controller-directed control was better suited to steering. %
Although controller-directed teleportation underperformed, controller-directed steering was broadly comparable to similar techniques in high-fidelity VR, and participants valued its flexibility and intuitiveness. These findings warrant further research into low-fidelity input devices and demonstrate that they can support effective locomotion in MVR, strengthening its potential as an accessible entry point to XR. %

\begin{acks}
We acknowledge the support of the Natural Sciences and Engineering Research Council of Canada (NSERC).
\end{acks}

\bibliographystyle{ACM-Reference-Format}
\bibliography{bibliography}

\end{document}